# THE WIGNER-VLASOV FORMALISM FOR TIME-DEPENDENT QUANTUM OSCILLATOR

**E.E. Perepelkin**[a,b,d*], **B.I. Sadovnikov**[a], **N.G. Inozemtseva**[b,c], **A.A. Korepanova**[a]

[a] *Faculty of Physics, Lomonosov Moscow State University, Moscow, 119991 Russia*
[b] *Moscow Technical University of Communications and Informatics, Moscow, 123423 Russia*
[c] *Dubna State University, Moscow region, Dubna,141980 Russia*
[d] *Joint Institute for Nuclear Research, Moscow region, Dubna,141980 Russia*
*Corresponding author: pevgeny@jinr.ru*

**Abstract**

This paper presents a comprehensive investigation of the problem of a harmonic oscillator with time-depending frequencies in the framework of the Vlasov theory and the Wigner function apparatus for quantum systems in the phase space. A new method is proposed to find an exact solution of this problem using a relation of the Vlasov equation chain with the Schrödinger equation and with the Moyal equation for the Wigner function.

A method of averaging the energy function over the Wigner function in the phase space can be used to obtain time-dependent energy spectrum for a quantum system. The Vlasov equation solution can be represented in the form of characteristics satisfying the Hill equation. A particular case of the Hill equation, namely the Mathieu equation with unstable solutions, has been considered in details.

An analysis of the dynamics of an unstable quantum system shows that the phase space square bounded with the Wigner function level line conserves in time, but the phase space square bounded with the energy function line increases. In this case the Vlasov equation characteristic is situated on the crosspoint of the Wigner function level line and the energy function line. This crosspoint moves in time with a trajectory that represents the unstable system dynamics. Each such trajectory has its own energy, and averaging these energies over the Wigner function results in time-dependent discreet energy spectrum for the whole system. An explicit expression has been obtained for the Wigner function of the 4th rank in the generalized phase space $\{x, p, \dot{p}, \ddot{p}\}$.

**Introduction**

The Schrödinger equation is the fundamental equation of quantum mechanics [1, 2]. As a rule, for an arbitrary potential $U^1$, the solution of the equation is sought by the numerical method [3, 4]. Exact solutions are known only for a small number of potentials $U^1$ [5-8]. The presence of a time-dependent potential $U^1(\vec{r}, t)$ significantly complicates obtaining an explicit form of the solution of the Schrödinger equation. Knowing the exact solution is important for several reasons, for example: methodical, theoretical and applied. An explicit solution allows for analyzing the behavior of a quantum system. When considering complex multicomponent systems, for example, in quantum chemistry and molecular dynamics, the numerical algorithm may contain encapsulations of exact solutions of individual parts of the system when it is modeled on the massive parallel processing of GPUs [9-11].

A wide class of physical systems has a time-dependent behavior. The oscillator model is a «zero» approximation in the description of a complex multicomponent dynamical system [35, 36]. The simplest time-dependent potential of an oscillator is

$$U^1(x,t) = \frac{m\Omega^2(t)x^2}{2}, \tag{i.1}$$

where the frequency $\Omega(t)$ is an arbitrary smooth function of time. Solutions of the Schrödinger equation with potential (i.1) was previously investigated in [37]. Exact solutions of the Schrödinger equation are of interest not only for quantum, but also for classical systems. As the Schrödinger equation is related to an infinite self-linking chain of Vlasov equations [12,13], it is possible to map solutions of some equations into solutions of other equations. Indeed, consider the first two equations from the infinite Vlasov chain for the distribution functions $f^1(\vec{r},t)$ and $f^{1,2}(\vec{r},\vec{v},t)$, respectively:

$$\frac{\partial f^1}{\partial t} + \text{div}_r \left[ f^1 \langle \vec{v} \rangle_1 \right] = 0, \tag{i.2}$$

$$\frac{\partial f^{1,2}}{\partial t} + \text{div}_r \left[ f^{1,2}\vec{v} \right] + \text{div}_v \left[ f^{1,2} \langle \dot{\vec{v}} \rangle_{1,2} \right] = 0, \tag{i.3}$$

where $\langle \vec{v} \rangle_1 (\vec{r},t)$ is the average vector field of the velocity of the probability flow, and $\langle \dot{\vec{v}} \rangle_{1,2}(\vec{r},\vec{v},t)$ is the average vector field of the flow of the accelerations of probabilities, satisfying the relations:

$$f^1\langle\vec{v}\rangle_1 = \int_{(\infty)} f^{1,2}\vec{v}\,d^3v, \qquad f^{1,2}\langle\dot{\vec{v}}\rangle_{1,2} = \int_{(\infty)} f^{1,2,3}\dot{\vec{v}}\,d^3\dot{v}, \qquad f^1 = \int_{(\infty)} f^{1,2}d^3v, \tag{i.4}$$

where $f^{1,2,3}(\vec{r},\vec{v},\dot{\vec{v}},t)$ is the distribution function satisfying the third Vlasov equation. If the probability density function $f^1(\vec{r},t)$ is positive, i.e. $f^1 = |\Psi^1(\vec{r},t)|^2 \geq 0$, $\Psi^1 \in \mathbb{C}$, and the vector field $\langle \vec{v} \rangle_1 (\vec{r},t)$ can be decomposed according to the Helmholtz theorem into potential $\nabla_r \Phi^1$ and vortex $\vec{A}$ components

$$\langle \vec{v} \rangle_1 (\vec{r},t) = \frac{\hbar}{2m}\nabla_r \Phi^1(\vec{r},t) - \frac{q}{m}\vec{A}(\vec{r},t), \tag{i.5}$$

then the first Vlasov equation (i.2) transforms into the Schrödinger equation for a scalar particle with a charge $q$ and mass $m$ in electromagnetic field [14, 15]:

$$i\hbar\frac{\partial}{\partial t}\Psi^1 = \frac{1}{2m}\left(\hat{p}-q\vec{A}\right)^2 \Psi^1 + U^1\Psi^1, \tag{i.6}$$

$$-\frac{\hbar}{2}\frac{\partial \Phi^1}{\partial t} = \frac{m}{2}|\langle\vec{v}\rangle_1|^2 + V \stackrel{\text{det}}{=} H, \tag{i.7}$$

$$V \stackrel{det}{=} U^1 + Q^1, \quad Q \stackrel{det}{=} -\frac{\hbar^2}{2m}\frac{\Delta_r |\Psi^1|}{|\Psi^1|}, \tag{i.8}$$

$$\frac{d}{dt}\langle \vec{v}\rangle_1 = \left(\frac{\partial}{\partial t} + \langle \vec{v}\rangle_1 \nabla_r\right)\langle \vec{v}\rangle_1 = \frac{q}{m}\left(\vec{E} + \langle \vec{v}\rangle_1 \times \vec{B}\right), \tag{i.9}$$

$$\vec{E} \stackrel{det}{=} -\frac{\partial}{\partial t}\vec{A} - \frac{1}{q}\nabla_r V, \quad \vec{B} \stackrel{det}{=} \operatorname{curl}_r \vec{A}, \tag{i.10}$$

where $\hat{p} \stackrel{det}{=} -i\hbar\nabla_r$ is a momentum operator. The scalar potential $\Phi$ is related to the phase $\varphi^1$ of the wave function $\Psi^1 = \pm\sqrt{f^1}\exp(i\varphi^1)$ by the relation $\Phi^1 = 2\varphi^1 + 2\pi k, k \in \mathbb{Z}$. Expression (i.7) corresponds to the Hamilton-Jacobi equation, in which $\hbar\varphi$ acts as an action. The potential V is the sum of the Schrödinger potential $U^1$ and the quantum potential $Q^1$ (i.8). The quantum potential $Q^1$ is used in the de Broglie-Bohm «pilot-wave» theory [16-18]. The equation of motion (i.9) contains the electromagnetic force with fields $\vec{E}, \vec{B}$ and is obtained directly from Hamilton-Jacobi equation (i.7).

Using the dynamic Vlasov-Moyal approximation for the kinematical average value $\langle \dot{\vec{v}}\rangle_{1,2}$ [19]

$$\langle \dot{v}_\mu\rangle_{1,2} = \sum_{l=0}^{+\infty}\frac{(-1)^{l+1}(\hbar/2)^{2l}}{m^{2l+1}(2l+1)!}\frac{\partial^{2l+1}U}{\partial x_\mu^{2l+1}}\frac{1}{f^{1,2}}\frac{\partial^{2l}f^{1,2}}{\partial v_\mu^{2l}}, \tag{i.11}$$

the second Vlasov equation (i.3) transforms into the Moyal equation [20] for the Wigner function $f^{1,2}(\vec{r},\vec{v},t) = mW(\vec{r},\vec{p},t)$

$$\frac{\partial W}{\partial t} + \frac{1}{m}\vec{p}\cdot\nabla_r W - \nabla_r U \cdot \nabla_p W = \sum_{l=1}^{+\infty}\frac{(-1)^l(\hbar/2)^{2l}}{(2l+1)!}U^1\left(\overleftarrow{\nabla}_r \cdot \overrightarrow{\nabla}_p\right)^{2l+1}W. \tag{i.12}$$

The Wigner function plays the role of a quasi-probability density for a quantum system in the phase space [21, 22]. Multiplication by the velocity $\vec{v}$ of the second Vlasov equation (i.3) and its subsequent integration over this space $\int\{\cdot\}\vec{v}d^3v$ gives an analog of equation (i.9) in the hydrodynamic approximation [12, 23]:

$$\frac{d}{dt}\langle v_\mu\rangle_1 = \left(\frac{\partial}{\partial t} + \langle v_\lambda\rangle_1 \frac{\partial}{\partial x_\lambda}\right)\langle v_\mu\rangle_1 = -\frac{1}{f^1}\frac{\partial P_{\mu\lambda}}{\partial x_\lambda} + \langle \dot{v}_\mu\rangle_1, \tag{i.13}$$

$$P_{\mu\lambda} \stackrel{det}{=} \int_{(\infty)} f^{1,2}\left(v_\mu - \langle v_\mu\rangle_1\right)\left(v_\lambda - \langle v_\lambda\rangle_1\right)d^3v,$$

where $P_{\mu\lambda}$ is a pressure tensor. It follows from a comparison of equations (i.9) and (i.13) that in the absence of the vortex field component $\vec{A}$ (i.5), the pressure tensor $P_{\mu\lambda}$ is related to the quantum pressure potential Q by the relation

$$\frac{1}{m}\frac{\partial Q^1}{\partial x_\mu} = \frac{1}{f^1}\frac{\partial P_{\mu\lambda}}{\partial x_\lambda}. \quad (i.14)$$

The use of expressions (i.5)-(i.14) makes it possible to obtain from solutions obtained for quantum systems solutions corresponding to classical mechanics, plasma physics, statistical physics, accelerator physics, field theory and continuum mechanics [24-26].

The purpose of this paper is to obtain an explicit expression for the exact solution of the Schrödinger equation (i.6) with potential (i.1) from classical physics (i.2), (i.3), (i.5), (i.7) and to consider its properties of dynamic stability in the phase space using the apparatus of the Wigner and Vlasov functions.

The paper has the following structure. In §1, we seek an exact time-dependent solution of the first Vlasov equation (i.2). By specially choosing the function $f_n^1(x,t)$, $n = 0,1...$, the probability velocity field $\langle v \rangle_{1|n}(x,t)$ is found that satisfies equations of motion (i.9) and (i.13). In §2, using expansion (i.5), the wave function $\Psi_n^1(x,t)$ is constructed that satisfies the Schrödinger equation (i.6) with potential (i.1). Knowing the wave function in §3, one can construct the Wigner function that satisfies the Moyal equation (i.12). The investigated Moyal equation admits solutions in the form of characteristics, which are the solution of the Hill equation [27]. As an example, a particular case of the Hill equation, the Mathieu equation [28], is considered. When certain parameters are chosen on the Ince-Strutt diagram, the Mathieu equation has unstable solutions, which are discussed in detail in §3. The time evolution of the energy spectrum of an unstable quantum system is studied. In the phase space, using the Wigner function, the process of unstable dynamics of phase trajectories is described. Further in §4 we consider the extension of the Wigner function to the phase space of higher kinematical values $\{x, p, \dot{p}, \ddot{p}\}$. For the ground state of the quantum system ($n = 0$), the Wigner function of the 4th rank $W^{1,2,3,4}(x, p, \dot{p}, \ddot{p}, t)$ was explicitly obtained, which satisfies the Moyal equation of 4-th rank [29]. The Appendix contains the proofs of the theorems and intermediate calculations.

**§1 Solution of the first Vlasov equation**

Consider a time-dependent probability density distribution of the form:

$$f_n^1(x,t) = \frac{1}{2^n n!}\frac{1}{\sqrt{2\pi}\sigma(t)}\exp\left(-\frac{x^2}{2\sigma^2(t)}\right)H_n^2\left(\frac{x}{\sqrt{2}\sigma(t)}\right), \quad n = 0,1,2..., \quad (1.1)$$

where $\sigma(t)$ is some known sufficiently smooth and positive function of time; $H_n$ are the Hermite polynomials (physical definition). Distribution (1.1) is normalized to unity for any instant of time.

**Theorem 1** *Let the function (1.1) satisfy the first Vlasov equation (i.2):*

$$\frac{\partial f_n^1}{\partial t} + \langle v \rangle_1 \frac{\partial f_n^1}{\partial x} + f_n^1 \frac{\partial \langle v \rangle_1}{\partial x} = 0, \quad (1.2)$$

*where $\langle v \rangle_1(x,t)$ is the average velocity of the probability flow. Then field $\langle v \rangle_1(x,t)$ satisfies the equation*

$$\frac{\partial \langle \tilde{v} \rangle_1}{\partial \tilde{x}} + 2 \langle \tilde{v} \rangle_1 \left[ -\tilde{x} + 2n \frac{H_{n-1}(\tilde{x})}{H_n(\tilde{x})} \right] = 2\sqrt{2} \left[ -\tilde{x}^2 + 2n \frac{H_{n-1}(\tilde{x})}{H_n(\tilde{x})} \tilde{x} + \frac{1}{2} \right] \dot{\sigma}, \quad (1.3)$$

*where* $\tilde{x}(x,t) \stackrel{det}{=} \frac{x}{\sqrt{2}\sigma(t)}$, $\langle v \rangle_1(x,t) = \langle v \rangle_1 \left( \tilde{x}\sqrt{2}\sigma(t), t \right) \stackrel{det}{=} \langle \tilde{v} \rangle_1 (\tilde{x},t)$; $\dot{\sigma}$ *is the time derivative of function* $\sigma(t)$. *In this case, the solution of equation (1.3) has the form:*

$$\langle v \rangle_{1|n}(x,t) = \frac{C}{H_n^2 \left( \frac{x}{\sqrt{2}\sigma(t)} \right)} \exp \left( \frac{x^2}{2\sigma(t)^2} \right) + \frac{\dot{\sigma}(t)}{\sigma(t)} x, \; n = 0,1,... \quad (1.4)$$

*where $C$ does not depend on $x$ coordinate.*

The proof of Theorem 1 is given in Appendix A.
Thus, expressions (1.1) and (1.4) determine the solution of the first Vlasov equation (1.2).

**Remark**

Note that the average probability flow (1.4) generally has poles at points $x_k^{(n)} : H_n \left( x_k^{(n)} \right) = 0$, $k = 1...n$. Function $\langle v \rangle_{1|0}(x,t)$ has no poles only for the ground state ($n=0$). It follows from expression (1.4) that for large values of $x$ the average probability velocity flow grows indefinitely, while the probability density (1.1) tends to zero. As a result, the probability current density $J_{1|n}(x,t) \stackrel{det}{=} f_n^1(x,t) \langle v \rangle_{1|n}(x,t)$ for a certain instant of time $t$ tends to $\frac{1}{2^n n!} \frac{C}{\sqrt{2\pi}\sigma(t)}$ at $x \to \pm\infty$. If we put an arbitrary value $C = 0$, then there will be no probability current at infinity. Velocity field $\langle v \rangle_{1|n}(x,t)$ will not depend on state number «$n$», i.e.

$$\langle v \rangle_1 (x,t) = \frac{\dot{\sigma}(t)}{\sigma(t)} x. \quad (1.5)$$

Since function $\sigma(t)$ is positive, the direction of velocity (1.5) is determined by derivative $\dot{\sigma}(t)$. Value $\sigma(t)$ corresponds to the standard deviation for distribution function (1.1). An increase of $\sigma(t)$ according to (1.5) will lead to probability waves from the center of the system ($x=0$) to its periphery $x \to \pm\infty$. A decrease of $\sigma(t)$ will cause the waves of probabilities to tend towards the center of the system. The first summand in expression (1.4) with multiplier $C \neq 0$ introduces an additional unidirectional flow from left to right ($C > 0$) or from right to left ($C < 0$) into the described process.

As an example, Fig. 1 (on the left) shows the distribution of velocity component $\langle v \rangle_{1|4}(x,t)$ (state number $n=4$) over the coordinate and time, and, on the right in Fig. 1, the corresponding probability density distribution $f_4^1(x,t)$ is shown. Without loss of generality, function $\sigma(t)$ was taken

$$\sigma(t) = \sigma_0 \left[ 1 + \sin^2(\varpi_0 t) \right], \qquad (1.6)$$

where $\sigma_0, \varpi_0$ are constant values. In Fig. 1 (on the left) the arrows show the directions of the velocity of probability flow $\langle v \rangle_{1|4}(x,t)$ at different points in time. The red color corresponds to a positive, and the blue color corresponds to a negative value of the velocity $\langle v \rangle_{1|4}(x,t)$ projection. On the right in Fig. 1, the arrows indicate the states of the system with maximum compression and expansion. The red color corresponds to the maximum, and the blue color – to the minimum value of probability density $f_4^1(x,t)$. At the moments of maximum compression/expansion (see Fig. 1 on the right) the average probability flow velocity is zero (see Fig. 1 on the left).

By virtue of the periodic law (1.6), the states of compression and expansion alternate. White wavy lines on the left in Fig. 1 show the location of the poles of function (1.4). Since the state with number $n = 4$ is considered, in Fig. 1 (left) there are four lines of zeros $x_k^{(4)}(t)$, $k = 1...4$ of the Hermitian polynomials.

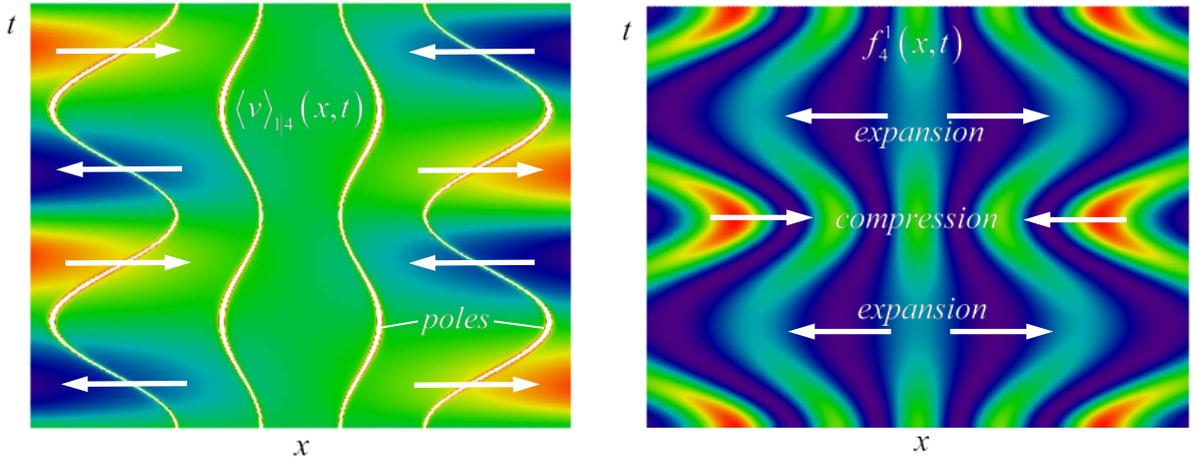

Fig. 1 Evolution of distributions $\langle v \rangle_{1|4}(x,t)$ and $f_4^1(x,t)$.

The presence of poles $x_k^{(n)}$, $k = 1...n$ leads to the partition of the domain of the function $\langle v \rangle_{1|n}(x,t)$ along the coordinate axis into intervals:

$$\left(-\infty, x_1^{(n)}\right), \left(x_1^{(n)}, x_2^{(n)}\right), ..., \left(x_{n-1}^{(n)}, x_n^{(n)}\right), \left(x_n^{(n)}, +\infty\right). \qquad (1.7)$$

As shown in [25, 30, 31], poles $x_k^{(n)}$ correspond to the poles of energy function $\langle \varepsilon \rangle_1(x,t)$

$$\langle \varepsilon \rangle(x,t) = \frac{1}{f^1(x,t)} \int_{-\infty}^{+\infty} f^{1,2}(x,v,t) \varepsilon(x,v,t) dv, \qquad (1.8)$$

$$\varepsilon(x,v,t) = \frac{mv^2}{2} + U^1(x,t), \qquad (1.9)$$

where $U^1(x,t)$ is the potential energy of the system from the Schrödinger equation (i.6), and distribution function $f^{1,2}(x,v,t)$ satisfies the second Vlasov equation (i.3) and corresponds to Wigner function $W(x,p,t)$, i.e. $f^{1,2}(x,v,t) = mW(x,mv,t)$. Poles $x_k^{(n)}$ correspond to infinite energy barriers that divide the region into «separate» oscillators with their own energy spectra. In each such interval, it is possible to build their own wave functions.

## §2 Wave function for the Schrodinger equation

Knowing the solution (1.1)/(1.4) of the first Vlasov equation (1.2), one can construct the corresponding solution of the Schrödinger equation [14, 24-26]. The average probability flow (1.4) is related to phase $\varphi^1(x,t)$ of wave function $\Psi^1(x,t)$ by the Helmholtz theorem (i.7):

$$\langle v \rangle_1 (x,t) = -\alpha \frac{\partial \Phi^1}{\partial x} = C \cdot H_n^{-2}\left(\frac{x}{\sqrt{2}\sigma(t)}\right) \exp\left(\frac{x^2}{2\sigma(t)^2}\right) + \frac{\dot{\sigma}(t)}{\sigma(t)} x, \qquad (2.1)$$

where $\Phi^1 = 2\varphi^1 + 2\pi j$, $j \in \mathbb{Z}$. According to (2.1), the phase is determined up to a value that does not depend on coordinate $E_n(t)$, therefore, integrating expression (2.1), we obtain:

$$\varphi_n^1(x,t) = -\frac{C}{2\alpha}\Theta_n(x,x_0,t) - \frac{\dot{\sigma}(t)}{4\alpha\sigma(t)}x^2 - \beta E_n(t), \qquad (2.2)$$

where

$$\Theta_n(x,x_0,t) \stackrel{\text{det}}{=} \sigma\sqrt{2} \int_{\frac{x_0}{\sigma\sqrt{2}}}^{\frac{x}{\sigma\sqrt{2}}} e^{\tilde{x}^2} H_n^{-2}(\tilde{x}) d\tilde{x}. \qquad (2.3)$$

Constant values $\alpha, \beta$ equal $\alpha = -\hbar/2m$, $\beta = 1/\hbar$. Point $x_0$ in integral (2.3) is chosen from the interval (1.7), in which phase (2.2) is considered. Taking into account expressions (1.1) and (2.2), wave function $\Psi_n^1(x,t)$ will take the form:

$$\Psi_n^1(x,t) = \text{sgn}(H_n) \sqrt{f_n^1(x,t)} e^{i\varphi_n^1(x,t)}, \qquad (2.4)$$

which satisfies the Schrödinger equation (i.5):

$$i\hbar \frac{\partial \Psi_n^1}{\partial t} = -\frac{\hbar^2}{2m} \frac{\partial^2 \Psi_n^1}{\partial x^2} + U^1 \Psi_n^1, \qquad (2.5)$$

where, according to the Hamilton-Jacobi equation (i.6), potential $U^1(x,t)$ has the form:

$$-\beta U^1(x,t) = \frac{\partial \varphi_n^1}{\partial t} + \alpha \frac{1}{\sqrt{f_n^1}} \frac{\partial^2 \sqrt{f_n^1}}{\partial x^2} - \alpha \left(\frac{\partial \varphi_n^1}{\partial x}\right)^2. \qquad (2.6)$$

Note that function sgn in expression (2.4) is necessary to preserve the differentiability property of the wave function at points $x_k^{(n)}$. By virtue of the remark from §1, from a physical point of view, it is interesting to consider the case when $C = 0$, i.e., the velocity field (1.5). Let us write expressions (2.4) and (2.6) for the case $C = 0$.

**Theorem 2** *Wave function*

$$\Psi_n^1(x,t) = \frac{1}{\sqrt{2^n n!}} \frac{1}{\sqrt{\sqrt{2\pi}\sigma(t)}} \exp\left(-\frac{x^2}{4\sigma^2(t)} - i\frac{\dot{\sigma}(t)}{4\alpha\sigma(t)} x^2 - i\beta E_n(t)\right) H_n\left(\frac{x}{\sqrt{2}\sigma(t)}\right), \quad (2.7)$$

*is a solution of the Schrödinger equation (2.5) with the potential*

$$U^1(x,t) = \frac{1}{4\alpha\beta\sigma(t)} \left[\ddot{\sigma}(t) - \frac{\alpha^2}{\sigma^3(t)}\right] x^2, \quad (2.8)$$

*wherein*

$$\dot{E}_n(t) \stackrel{det}{=} -\frac{\alpha}{\beta\sigma^2(t)}\left(n + \frac{1}{2}\right). \quad (2.9)$$

The proof of Theorem 2 is given in Appendix B.

Note that function $E_n(t)$, due to its arbitrariness, is determined from differential equation (2.9). Thus, function (2.7) is a solution of the Schrödinger equation (2.5) with time-dependent potential (2.8). The eigenenergy spectrum depends on time and is related to equation (2.9).

In the particular case of time-independent distribution (1.1), for example, for

$$\sigma_0^2 = \frac{|\alpha|}{\omega_0} = \frac{\hbar}{2m\omega_0}, \quad (2.10)$$

expressions (2.7)-(2.9) transform into the known relations for a quantum harmonic oscillator:

$$\Psi_{n,\sigma=const}^1(x,t) = \frac{1}{\sqrt{2^n n!}} \left(\frac{m\omega_0}{\pi\hbar}\right)^{1/4} \exp\left(-\frac{m\omega_0 x^2}{2\hbar} - i\frac{E_n}{\hbar} t\right) H_n\left(x\sqrt{\frac{m\omega_0}{\hbar}}\right), \quad (2.11)$$

$$U_{\sigma=const}^1(x) = \frac{m\omega_0^2 x^2}{2}, \quad E_{n,\sigma=const}(t) = E_n t, \quad E_n = \hbar\omega_0\left(n + \frac{1}{2}\right).$$

Note that finding the spectrum of energies $E_n(t)$ for the time-dependent case is possible using the Wigner function, by averaging energy function (1.9) over the entire phase space. This procedure is considered in the next section.

The expression for potential (2.8) can be rewritten as:

$$U^1(x,t) = \frac{m\Omega^2(t)x^2}{2}, \quad \Omega^2(t) \stackrel{det}{=} \frac{1}{\sigma}\left(\frac{\alpha^2}{\sigma^3} - \ddot{\sigma}\right). \quad (2.12)$$

Substituting expressions (1.5) and (2.12) into equation of motion (i.9) gives the correct identity:

$$\frac{d}{dt}\langle v\rangle_{1|n} = \frac{\ddot{\sigma}}{\sigma} x = -\frac{1}{m}\frac{\partial}{\partial x}(U + Q_n), \quad Q_n(x,t) = -\frac{\hbar^2}{8m\sigma^4}\left[x^2 - 2\sigma^2(1+2n)\right], \quad (2.13)$$

where the expression for the quantum potential is obtained in Appendix B. Substitution of (1.5), (2.2), (2.12) and expression (2.13) for $Q_n(x,t)$ into the Hamilton-Jacobi equation (i.7) leads to equation (2.9).

The condition $\Omega(t)=0$ is an analogue of the separatrix equation for function $\sigma(t)$ (see Appendix B):

$$\frac{c_1^2}{\alpha^2}\sigma^2 - \frac{c_1^4}{\alpha^2}(t \pm c_2)^2 = 1, \tag{2.14}$$

where $c_1, c_2$ are constant values. Equation (2.14) defines a set of hyperbolas. If function $\sigma(t)$ satisfies condition (2.14), then potential $U^1(x,t)$ degenerates and the quantum system corresponds to the model of a free particle. Note that, at $\Omega^2(t) > 0$, force $-m\Omega^2(t)x$ acts on the particle, but, at $\Omega^2(t) < 0$ (imaginary value $\Omega(t)$), external force of the opposite sign $+m\Omega^2(t)x$ arises, which again acts on the particle.

### §3 Wigner function

To analyze the system in the phase space, we find the Wigner function of quantum state $n$ [21, 22]: [21, 22]:

$$W_n(x,p,t) = \frac{1}{2\pi\hbar} \int_{-\infty}^{+\infty} \overline{\Psi}_n^1\left(x - \frac{s}{2}, t\right) \Psi_n^1\left(x + \frac{s}{2}, t\right) e^{-i\frac{ps}{\hbar}} ds. \tag{3.1}$$

**Theorem 3** *The Wigner function (3.1) of the quantum system described by wave function (2.7) has the form:*

$$W_n(x,p,t) = \frac{(-1)^n}{\pi\hbar} e^{-\varepsilon(x,p,t)} L_n(2\varepsilon(x,p,t)), \tag{3.2}$$

*where*

$$\varepsilon(x,p,t) \stackrel{\text{det}}{=} \kappa^2(t)x^2 + \left[\sigma(t)\dot{\sigma}(t)\frac{x}{\alpha}\kappa(t) + \frac{p}{\kappa(t)\hbar}\right]^2, \quad \kappa(t) \stackrel{\text{det}}{=} \frac{1}{\sigma(t)\sqrt{2}}. \tag{3.3}$$

The proof of Theorem 3 is given in Appendix C.

Note that, according to expression (i.4), the average velocity (1.5) satisfies the relation

$$\langle v \rangle_{1|n}(x,t) = \frac{m}{f_n^1} \int_{-\infty}^{+\infty} W_n(x,mv,t)v dv = \frac{\dot{\sigma}}{\sigma}x, \tag{3.4}$$

and the quantum pressure force (i.13), (i.14) has the form:

$$-\frac{1}{f_n^1}\frac{\partial P_{11}}{\partial x} = -\frac{m}{f_n^1}\frac{\partial}{\partial x}\int_{-\infty}^{+\infty} W_n(x,mv,t)\left[v - \langle v \rangle_{1|n}(x,t)\right]^2 dv = \frac{\alpha^2}{\sigma^4}x, \tag{3.5}$$

wherein value $\langle \dot{v}_\mu \rangle_1$ in equation of motion (i.13) according to the Vlasov-Moyal approximation (i.11) for potential $U^1$ (2.12) is represented as $\langle \dot{v} \rangle_{1|n} = -\Omega^2(t)x$.

The time-dependent Wigner function (3.2) transforms into known the Wigner function of a quantum harmonic oscillator under condition (2.10)

$$\varepsilon(x,p) = \frac{2}{\hbar\omega_0}\left(\frac{p^2}{2m} + \frac{m\omega_0^2 x^2}{2}\right). \tag{3.6}$$

In contrast to the time-independent case (3.6), phase trajectories (3.3) will be inclined ellipses changing their principal semi-axes with time

$$\left(1 + \frac{\sigma^2\dot{\sigma}^2}{\alpha^2}\right)\kappa^2 x^2 + 2\frac{\sigma\dot{\sigma}}{\alpha\hbar}xp + \frac{p^2}{\kappa^2\hbar^2} = const. \tag{3.7}$$

Fig. 2 shows the evolution of the Wigner function (3.2) for the state with number $n=2$ at instant of time (1.6) $t = 0, \tau, 2\tau, 3\tau$, where $\tau = \pi/4\varpi_0$. The red color corresponds to the maximum, and the purple color corresponds to the minimum value of the function. For the state with number $n=2$, the minimum value is negative, which indicates the quasi-probabilistic nature of the Wigner function. The red dotted line in Fig. 2 shows phase ellipses (3.7), and the red solid line shows their main axes.

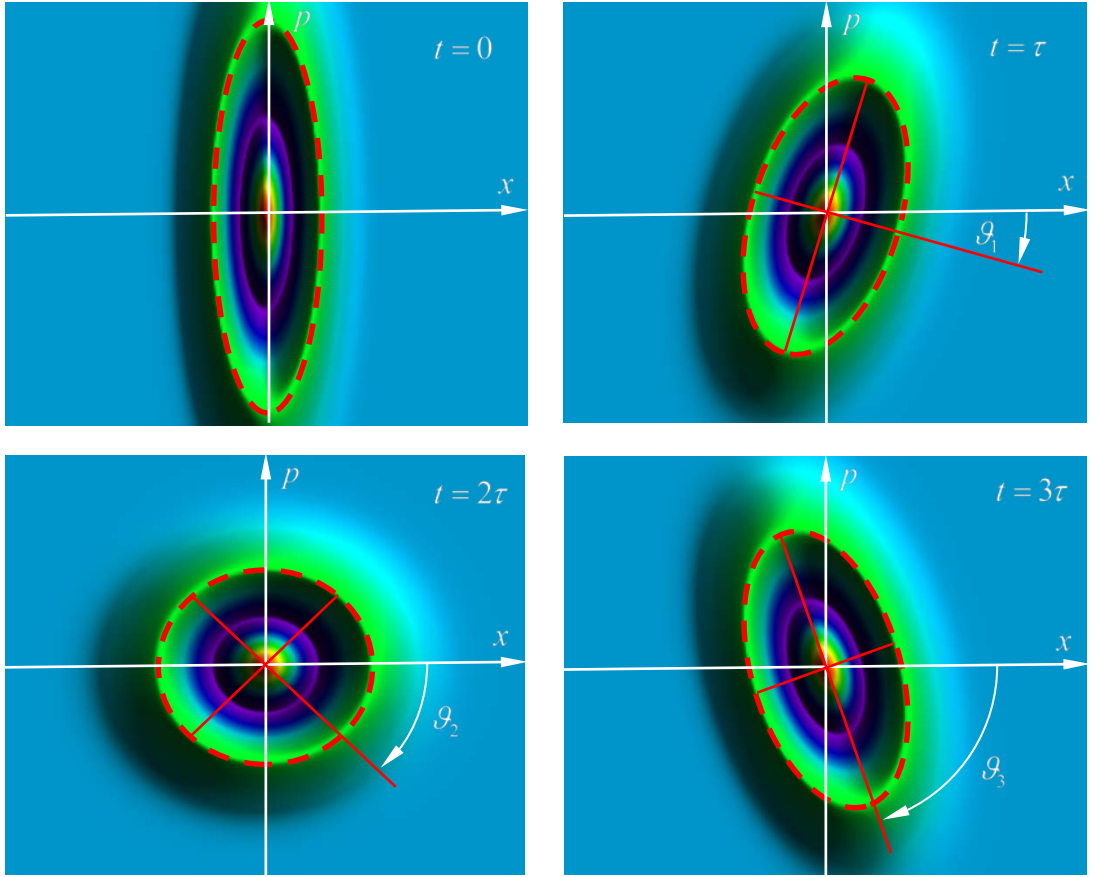

Fig. 2 Evolution of the Wigner function $W_4(x,p,t)$.

Distribution $W_2(x,p,t)$ rotates in the phase plane with respect to the origin. Angle of rotation $\vartheta(t)$ (see Fig. 2) according to (3.7) is determined by expression $\text{tg}(2\vartheta) = \dfrac{2\alpha}{\sigma\dot{\sigma}} = -\dfrac{\hbar}{m\sigma\dot{\sigma}}$ (see Appendix C). The values of the angles of rotation in Fig. 2 are found by the formula $\vartheta_j = \vartheta(j\tau)$, $j = 0...3$. In addition to the slope of phase ellipse (3.7), there is a change in the values of its principal axes (see Fig. 2). Despite the change in the shape of the phase ellipse, its area remains constant over time. For example, for the case of a singular right-hand side of equation (3.7), the area of the phase region of the ellipse is $\pi$ (see Appendix C). This area conservation is valid for all concentric ellipses that form the contour lines of the Wigner function (3.2). Since the regions of negative values of the Wigner function are also located between such ellipses, therefore, such regions will be preserved, which is consistent with the results of [23].

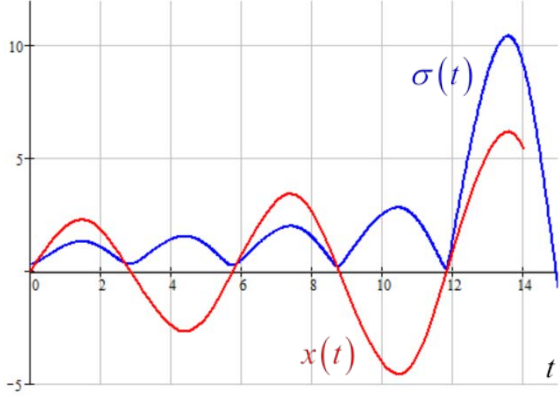

Fig. 3 Behavior of functions $x(t), \sigma(t)$ for the Mathieu equation.

On the one hand, the conservation of the phase area is in accordance with the Liouville theorem, written for the classical probability density function, which does not have negative values. On the other hand, for the Wigner function, there is a quantum generalization of the Liouville equation in the form of the Moyal equation (i.12). The quadratic dependence of the potential along the coordinate (2.12) leads to derivatives $\dfrac{\partial^{2k+1} U^1}{\partial x^{2k+1}}$ be equal to zero at $k > 0$. Therefore, the Moyal equation (i.12) for the Wigner function (3.2) takes the form:

$$\left[\frac{\partial}{\partial t} + \frac{p}{m}\frac{\partial}{\partial x} - \frac{\partial}{\partial x}U^1(x,t)\frac{\partial}{\partial p}\right]W_n(x,p,t) = 0. \qquad (3.8)$$

The form of equation (3.8) coincides with the form of the Liouville equation for the classical probability density. Direct substitution of expression (3.2) into equation (3.8) gives the characteristic equation:

$$\frac{d\varepsilon}{dt} = \frac{\partial \varepsilon}{\partial t} + \frac{p}{m}\frac{\partial \varepsilon}{\partial x} - m\Omega^2 x \frac{\partial \varepsilon}{\partial p} = 0, \qquad (3.9)$$

$$dt = m\frac{dx}{p} = -\frac{dp}{m\Omega^2 x},$$

$$\ddot{x} + \Omega^2(t)x = 0, \qquad (3.10)$$

which function (3.3)/(3.7) satisfies (see Appendix C). On the one hand, equation (3.10) is obtained from equation (3.9), and on the other hand, it is a natural consequence of Newton's second law with potential (2.12).

Equation (3.10) is well known in physics as the Hill equation [27]. A special case of the Hill equation is the Mathieu equation [28], in which $\Omega^2(t) = a - 2g\cos(2t)$, where $a, g \in \mathbb{R}$ are given parameters

$$\sigma^3\ddot{\sigma} + \sigma^4\left[a - 2g\cos(2t)\right] - \alpha^2 = 0. \tag{3.11}$$

By solving equation (3.11) one can find function $\sigma(t)$ transforming equation (3.10) into the Mathieu equation. Fig. 3 (blue color) shows solution $\sigma(t)$ of equation (3.11) with parameters $a = 1$ and $g = 0.2$. Fig. 3 shows that graph $\sigma(t)$ of the function increases its amplitude, which at the end has a strong surge.

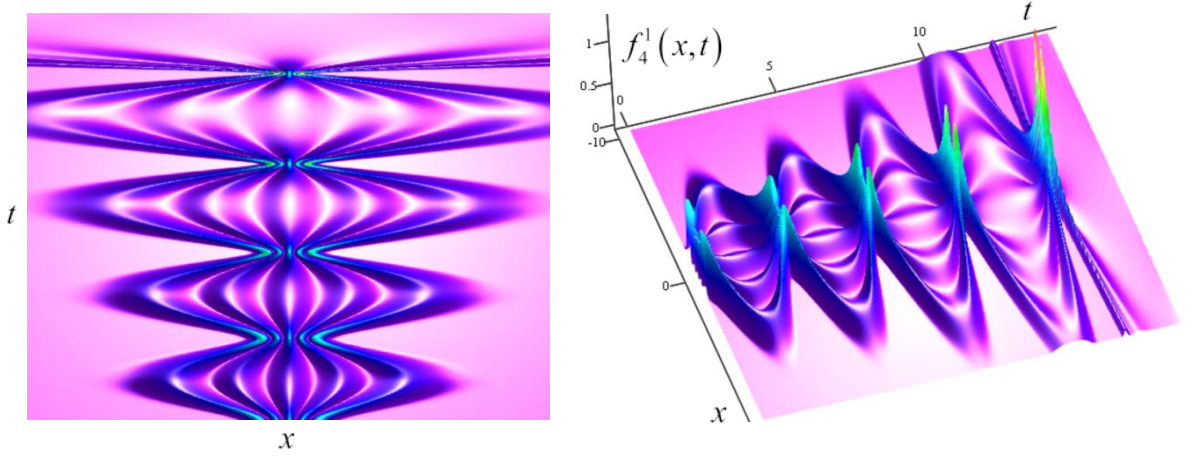

Fig. 4 Evolution of distribution $f_4^1(x,t)$.

The red color in Fig. 3 shows solution $x(t)$ of the Mathieu equation (3.10) corresponding to function $\sigma(t)$. The amplitude graph of solution $x(t)$ (red in Fig. 3) also grows. As is known, the Mathieu equation can have stable and unstable solutions $x(t)$ depending on parameters $(a, g)$. By choosing $(a, g)$ parameters on the Ince-Strutt diagram, one can obtain stable or unstable solutions of the Mathieu equation. Substituting the obtained dependence $\sigma(t)$ into the expressions for average probability flow rate $\langle v \rangle_{1|n}(x,t)$ (1.5), wave function $\Psi_n^1(x,t)$ (2.7), potential $U^1(x,t)$ (2.8)/(2.12), probability density distribution function $f_n^1(x,t)$ (1.1), energy spectrum $\mathrm{E}_n(t)$, it is possible to construct the corresponding distributions for each number $n$ of the quantum state of the system.

Fig. 4 shows the evolution of probability density distribution $f_4^1(x,t)$ for the quantum state with number $n = 4$. On the left in Fig. 4, a top view of distribution $f_4^1(x,t)$ is shown. The horizontal axis corresponds to coordinate $x$, and the vertical axis corresponds to time $t$. On the right in Fig. 4, an isometric view of distribution $f_4^1(x,t)$ is shown. On Fig. 4, it can be seen that the amplitude of the distribution fluctuations grows with time in accordance with the graphs in Fig. 3.

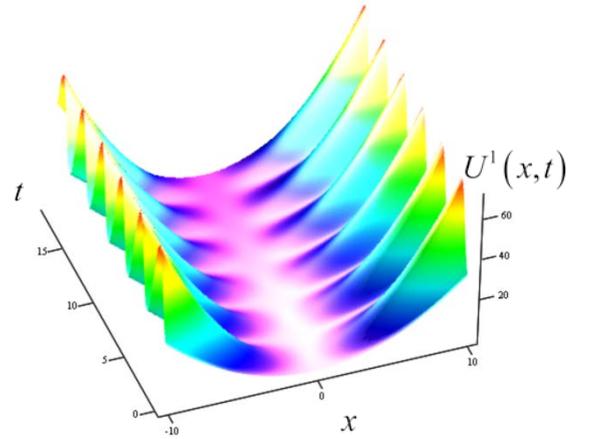

Fig. 5 Evolution of potential $U^1(x,t)$.

Potential energy $U^1(x,t)$ in accordance with expressions (2.12) and (3.11) has the form:

$$U^1(x,t) = \frac{mx^2}{2}\left[a - 2g\cos(2t)\right]. \tag{3.12}$$

The graph of function (3.12), shown in Fig. 5 is well known in accelerator physics. Distribution (3.12) (see Fig. 5) is used to transport a beam of charged particles through a magneto-optical structure with focusing (F) and defocusing (D) quadrupole magnetic lenses.

Let us find the energy spectrum of the system using the expression for the Wigner function (3.2).

$$\mathrm{E}_n(t) = \langle \mathcal{E} \rangle_0(t) = \int_{-\infty}^{+\infty}\int_{-\infty}^{+\infty} W_n(x,p,t)\mathcal{E}(x,p,t)\,dxdp, \tag{3.13}$$

$$\mathcal{E}(x,p,t) = \frac{p^2}{2m} + \frac{m\Omega^2(t)x^2}{2}. \tag{3.14}$$

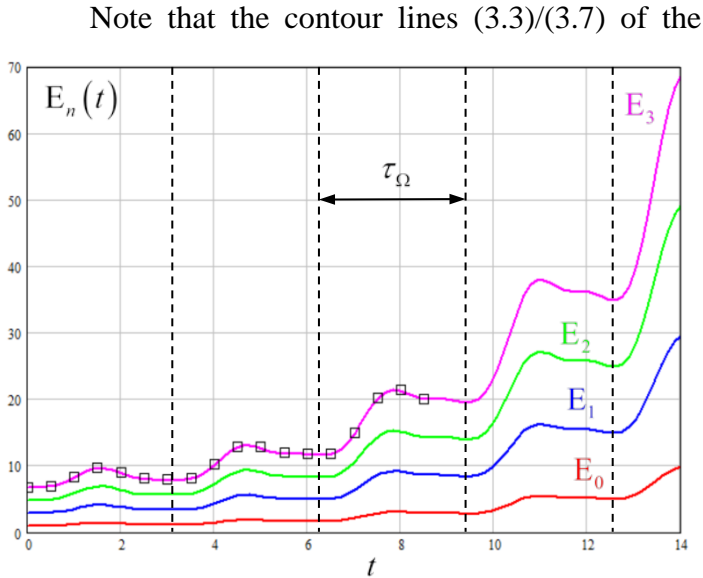

Fig. 6 Evolution of the spectrum of energy $\mathrm{E}_n(t)$

Note that the contour lines (3.3)/(3.7) of the Wigner function (3.2) generally do not coincide with trajectories (3.14) $\mathcal{E}(x,p,t) = const$. This is a significant difference from a time-independent quantum harmonic oscillator. Fig. 6 shows the time evolution of eigenenergy spectrum (3.13) of the system. As can be seen from Fig. 6, the energy spectrum has periodic bursts caused by the periodic behavior of the potential (3.12) (see Fig. 5). By analogy with the time-independent ($\Omega(t) = \omega_0 = const$) quantum harmonic oscillator in Fig. 6, an equidistant arrangement of energy levels is observed at each instant of time. The spectral energies increase with time (see Fig. 6) due to the instability of the system under consideration. An analogous process is observed for distribution function $f_4^1(x,t)$ in Fig. 4 and for particle trajectory $x(t)$ in Fig. 3. The vertical dotted lines in Fig. 6 mark periods $\tau_\Omega$ of frequency function $\Omega(0) = \Omega(\tau_\Omega k)$, $k \in \mathbb{N}$. At each instant of time $t = \tau_\Omega k$, the form of potential energy $U^1(x,0) = U^1(x,\tau_\Omega k)$ (3.12) is the same, but the spectrum of energy $\mathrm{E}_n(\tau_\Omega(k-1)) \neq \mathrm{E}_n(\tau_\Omega k)$ is different. This difference is caused by the energy «pumping» of the quantum system due to its instability.

Let us consider in more detail the relationship and meaning of characteristics $\mathcal{E}(x,p,t)$ (3.14) and $\varepsilon(x,p,t)$ (3.3)/(3.6). Fig. 7 demonstrates the graphs of functions $\varepsilon(x,p,t) = const$ (green) and $\mathcal{E}(x,p,t) = const$ (blue) at different instants of time $t_k = \{0, \tau, 2\tau, 3\tau, 5\tau\}$, $k = 0...4$. The red color in Fig. 7 shows the phase trajectory $\{x(t), p(t)\}$ of the solution of the Mathieu equation (3.10) with potential (3.12). The solution of the Mathieu equation is unstable, which leads to a phase trajectory in the form of an unwinding spiral (see Fig. 7). The motion of a particle along the phase trajectory is shown as a ball (magenta color) with a vector indicating the direction of its motion (see Fig. 7). The length of the vector is proportional to energy $\mathcal{E}(x,p,t)$ of the particle. The meaning of energies $\mathcal{E}_k = \mathcal{E}(x(t_k), p(t_k), t_k)$.

Expression $\mathcal{E} = const$ defines an ellipse in canonical form (3.14) in phase plane $(x, p)$. The principal semi-axes change with time due to two factors: frequency $\Omega(t)$ (3.14) and energy increase $\mathcal{E}(x, p, t)$. The area of the ellipse of constant $\mathcal{E} = const$ changes in proportion to the energy of the particle. Equation $\varepsilon = const$ defines ellipse (3.3)/(3.5), which rotates in the phase plane around the origin (see Fig. 2, 7), its semi-axes also change, but the area remains unchanged.

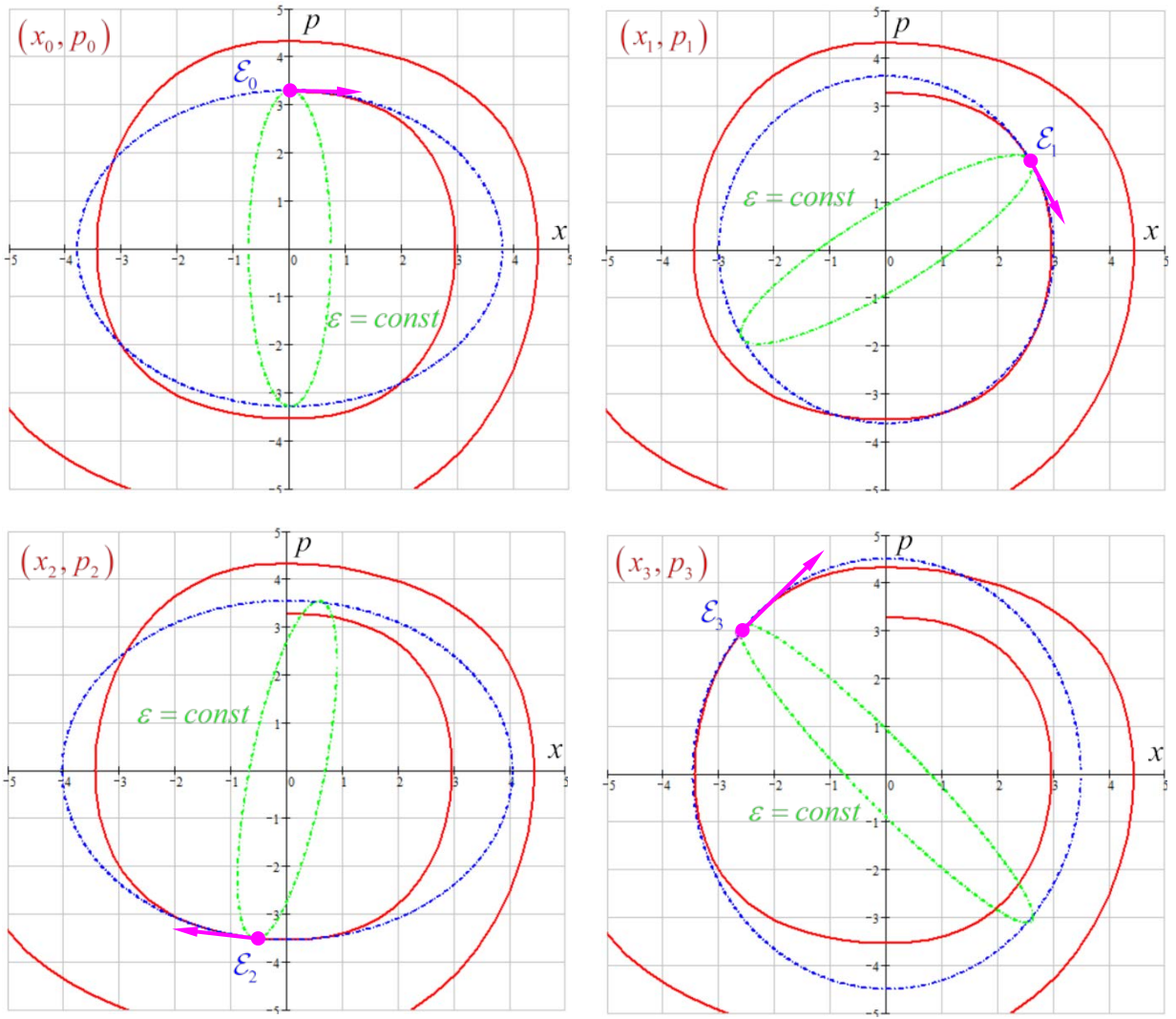

Fig. 7 Non-stable motion of a particle along the phase trajectory of the Wigner function

Note that both functions $\mathcal{E}(x, p, t)$ (3.14) and $\varepsilon(x, p, t)$ (3.3)/(3.6) are characteristics (see expressions (3.9) and (3.10)) of the Moyal/Liouville equation (3.8). On the one hand, the particle must move according to equation of motion (3.10), i.e., be on ellipse $\mathcal{E} = const$. On the other hand, according to the Liouville theorem, it is likely to be on phase trajectory $\varepsilon = const$. On line $\varepsilon = const$, in accordance with expression (3.2), the Wigner function is constant $W_n(\varepsilon = const) = const$. The described situation is possible if the particle is positioned on the phase plane at the intersection of two ellipses (3.14) and (3.3), which is shown in Fig. 7.

If at the initial moment of time ($t_0 = 0$) we calculate integral (3.13) for the quantum state with number $n$, then we can find $\mathrm{E}_n(0)$. Knowing $\mathrm{E}_n(0)$, we can determine the value of initial momentum $p_0^{(n)} = \sqrt{2m\mathrm{E}_n(0)}$ (coordinate $x_0^{(n)} = 0$ see Fig. 7). Values $\left(x_0^{(n)}, p_0^{(n)}\right)$ determine the Cauchy problem for the Mathieu equation (3.10), and, consequently, its solution in the form of phase trajectory $\{x^{(n)}(t), p^{(n)}(t)\}$. In fig. 7 (red color), the phase trajectory is constructed for the quantum state with number $n = 2$, i.e. $\{x^{(2)}(t), p^{(2)}(t)\}$. Knowing the phase trajectory makes it possible to use formula (3.14) instead of calculating integral (3.13) for each moment of time:

$$\mathrm{E}_n(t) = \mathcal{E}\left(x^{(n)}(t), p^{(n)}(t), t\right) = \frac{1}{2m}\left[p^{(n)}(t)\right]^2 + \frac{m\Omega^2(t)}{2}\left[x^{(n)}(t)\right]^2. \qquad (3.15)$$

The described procedure is applicable to each number $n$ of the quantum state of the system. The graphs of the evolution of the spectrum of energy $\mathrm{E}_n(t)$ eigenvalues shown in Fig. 6 are built exactly according to formula (3.15), and the dots in the form of squares mark the results obtained through integral (3.13). As seen in Fig. 6, the results obtained by both methods are the same. Thus, knowing the Wigner function in combination with the equations of motion makes it possible to find in a simple way the evolution of the energy spectrum of a time-dependent quantum system.

Note that in the case of a time-independent ($\Omega(t) = const$) quantum harmonic oscillator, ellipses $\mathcal{E} = const$ and $\varepsilon = const$ coincide, the system becomes stable, and the energy does not change. As noted above, in the time-dependent case, the stability of the system is determined by the choice of function $\Omega(t)$ in Hill's equation (3.10).

## §4 High kinematical solution

The Moyal equation for the Wigner function is a special case of the second Vlasov equation (i.3) for distribution function $f^{1,2}(\vec{r}, \vec{v}, t)$ with the Vlasov-Moyal approximation (i.11) for average acceleration flow $\langle\dot{\vec{v}}\rangle_{1,2}$. According to the second Vlasov equation, one can construct the Schrödinger equation of the second rank [15] using representation $f^{1,2} = |\Psi^{1,2}|^2 \geq 0$ and the expansion according to the Helmholtz theorem with the Vlasov-Moyal approximation:

$$\langle\dot{v}\rangle_{1,2}(x,v,t) = -\frac{1}{m}\frac{\partial}{\partial x}U^1(x,t) = -\frac{1}{m}\left(\frac{\hbar^2}{4m\sigma^4} - m\frac{\ddot{\sigma}}{\sigma}\right)x = -\alpha_2\frac{\partial}{\partial v}\Phi^{1,2}(x,v,t), \qquad (4.1)$$

$$\Phi^{1,2}(x,v,t) = -\left(\frac{\hbar^2}{2m\sigma^4} - 2m\frac{\ddot{\sigma}}{\sigma}\right)\frac{xv}{\hbar_2} + h(x,t), \qquad (4.2)$$

where $\alpha_2 = -\frac{\hbar_2}{2m}$ is a constant value, and $h(x,t)$ is some arbitrary function. Scalar potential (4.2) is related to phase $\varphi^{1,2}$ of wave function $\Psi^{1,2}$ by the relation $\Phi^{1,2} = 2\varphi^{1,2} + 2\pi k, k \in \mathbb{Z}$. Since, according to the Hudson theorem [34], the Wigner function (3.2) is positive only for the ground state ($n = 0$) of a quantum system, then wave function $\Psi^{1,2}$ of the second rank can be written as:

$$\Psi^{1,2}(x,v,t) = \sqrt{W_0(x,p,t)}e^{i\varphi^{1,2}(x,v,t)} = \frac{1}{\sqrt{\pi\hbar}}e^{-\frac{x^2}{4\sigma^2} - \frac{m^2}{\hbar^2}(\dot{\sigma}x - \sigma v)^2 - i\left(\frac{\hbar^2}{2m\sigma^4} - 2m\frac{\ddot{\sigma}}{\sigma}\right)\frac{xv}{2\hbar_2} + \frac{i}{2}h(x,t)}. \qquad (4.3)$$

Note that under condition (2.10) the wave function (4.3) takes the form:

$$\Psi^{1,2}(x,v,t) = \frac{1}{\sqrt{\pi\hbar}}\exp\left[-\frac{1}{\hbar\omega_0}\left(\frac{mv^2}{2} + \frac{m\omega_0^2 x^2}{2}\right) - i\left(\frac{m\omega_0^2}{\hbar_2}xv + \frac{E^{1,2}(t)}{\hbar_2}\right)\right], \qquad (4.4)$$

where $h(x,t) = -2\frac{E^{1,2}(t)}{\hbar_2}$. Function (4.3) coincides with the known solution of the Schrödinger equation of the second rank for a harmonic oscillator [32]. In case (4.4), function $E^{1,2}(t) = \mathrm{E}t$, $\mathrm{E} = \frac{\hbar_2\omega_0}{2}$.

**Theorem 4** *Wave function (4.4) is a solution of the second rank Schrödinger equation [15]:*

$$i\hbar_2\partial_1\Psi^{1,2} = \hat{\mathrm{H}}_{1,2}\Psi^{1,2}, \qquad (4.5)$$

$$i\hbar_2\left(\frac{\partial}{\partial t} + v\frac{\partial}{\partial x}\right)\Psi^{1,2} = -\frac{\hbar_2^2}{2m}\frac{\partial^2}{\partial v^2}\Psi^{1,2} + U^{1,2}\Psi^{1,2},$$

*with potential*

$$U^{1,2}(x,v,t) = \left(\frac{\hbar^2}{4m\sigma^4} - m\frac{\ddot{\sigma}}{\sigma} + 2\frac{\hbar_2^2 m^3\sigma^4}{\hbar^4}\right)v^2 - \left(\frac{\hbar^2\dot{\sigma}}{m\sigma^5} + m\frac{\ddot{\sigma}\sigma - \ddot{\sigma}\dot{\sigma}}{\sigma^2} + 4\frac{\hbar_2^2 m^3}{\hbar^4}\sigma^3\dot{\sigma}\right)xv + \qquad (4.6)$$

$$+ \left[2\frac{\hbar_2^2 m^3}{\hbar^4}\sigma^2\dot{\sigma}^2 - \left(\frac{\hbar^2}{4m\sqrt{2m}\sigma^4} - \frac{\ddot{\sigma}}{\sigma}\sqrt{\frac{m}{2}}\right)^2\right]x^2,$$

*where value* $\dot{E}^{1,2}(t) = \frac{m\hbar_2^2}{\hbar^2}\sigma^2(t)$.

The proof of Theorem 4 is given in Appendix D.

Potential (4.6), phase (4.2) and acceleration field (4.1) satisfy the Hamilton-Jacobi equation of the second rank [15]:

$$-\hbar_2\frac{\partial\varphi^{1,2}}{\partial t} = \frac{m}{2}\left|\langle\dot{v}\rangle_{1,2}\right|^2 + \mathrm{V}^{1,2} = \mathrm{H}^{1,2}, \qquad (4.7)$$

$$\mathrm{V}^{1,2} = U^{1,2} + \mathrm{Q}^{1,2}, \quad \mathrm{Q}^{1,2} = -\frac{\hbar_2^2}{2m}\frac{1}{|\Psi^{1,2}|}\frac{\partial^2}{\partial v^2}|\Psi^{1,2}|, \qquad (4.8)$$

where $Q^{1,2}$ is quantum potential of the second rank, which is an extended analogue of the quantum potential of the first rank $Q^1 = -\frac{\hbar^2}{2m}\frac{1}{|\Psi^1|}\frac{\partial^2}{\partial x^2}|\Psi^1|$ from the de Broglie-Bohm «wave-pilot» theory. Values $\hbar_2 \varphi^{1,2}$ and $H^{1,2}$ are, respectively, an action and the Hamiltonian function of the second rank. In the case under consideration in (4.1), the Hamiltonian function $H^{1,2}$ is related to the Lagrangian function $L^{1,2}$ through the Legendre transformation $H^{1,2} + L^{1,2} = \langle \dot{p} \rangle_{1,2} \langle \dot{v} \rangle_{1,2}$ [15].

In the time-independent case (2.10), expression (4.6) for potential $U^{1,2}(x,v,t)$ transforms [32] into potential $U^{1,2}(x,v)$ of a harmonic oscillator in the phase space:

$$U^{1,2}(x,v) = m\omega_0^2\left(1 + \frac{\hbar_2^2}{2\hbar^2\omega_0^4}\right)v^2 - \frac{m\omega_0^4}{2}x^2. \quad (4.9)$$

Using the expression for wave function $\Psi^{1,2}$ of the second rank (4.3), we find the Wigner function of the fourth rank $W^{1,2,3,4}(x,p,\dot{p},\ddot{p},t)$ [29].

$$W^{1,2,3,4}(x,p,\dot{p},\ddot{p}) = \frac{1}{(2\pi\hbar_2)^2}\int_{-\infty}^{+\infty}\int_{-\infty}^{+\infty}\overline{\Psi}^{1,2}(x',v',t)\Psi^{1,2}(x'',v'',t)\exp\left(i\frac{s_1\ddot{p} - s_2\dot{p}}{\hbar_2}\right)ds_1 ds_2, \quad (4.10)$$

where

$$x' = x - \frac{s_1}{2},\ v' = v - \frac{s_2}{2},\ x'' = x + \frac{s_1}{2},\ v'' = v + \frac{s_2}{2}.$$

**Theorem 5** *The Wigner function of the fourth rank (4.10) corresponds to the wave function of the second rank (4.3) and it has the form:*

$$W^{1,2,3,4}(x,p,\dot{p},\ddot{p},t) = \frac{1}{(\pi\hbar_2)^2}\exp\left[-\frac{x^2}{2\sigma^2} - \frac{2}{\hbar^2}(\dot{\sigma}mx - \sigma p)^2\right]\times$$

$$\times\exp\left\{-\frac{2}{m^2\hbar_2^2}\left(\left[\sigma(m\ddot{p} - \eta p) - m\dot{\sigma}(\dot{p} + \eta x)\right]^2 - \frac{\hbar^2}{4\sigma^2}(\dot{p} + \eta x)^2\right)\right\}, \quad (4.11)$$

*where*

$$\eta(t) \stackrel{\text{det}}{=} \frac{\hbar^2 - 4m^2\sigma^3\ddot{\sigma}}{4m\sigma^4}.$$

The proof of Theorem 5 is given in Appendix D.

In the time-independent case (2.10), expression (4.11) transforms into the function obtained in [29] for a harmonic oscillator:

$$W^{1,2,3,4}(x,p,\dot{p},\ddot{p}) = \frac{1}{(\pi\hbar_2)^2}\exp\left\{-\frac{2}{\hbar\omega_0}\left[\frac{p^2}{2m} + \frac{m\omega_0^2 x^2}{2} + \frac{(\dot{p} + m\omega_0^2 x)^2}{2m\omega_0^2} + \frac{(\ddot{p} - \omega_0^2 p)^2}{2m\omega_0^4}\right]\right\}. \quad (4.12)$$

According to [29], the Wigner function of the 4th rank (4.10) satisfies the Moyal $\Psi$-equation:

$$\partial_{1,2,3}W^{1,2,3,4} = \left(\frac{\partial}{\partial t} + \frac{p}{m}\frac{\partial}{\partial x} + \dot{p}\frac{\partial}{\partial p} + \ddot{p}\frac{\partial}{\partial \ddot{p}}\right)W^{1,2,3,4} =$$
$$= \sum_{l=0}^{+\infty}\sum_{n=0}^{2l+1}\frac{(-1)^{n+l}(\hbar_2/2m)^{2l}}{m^{n-1}n!(2l-n+1)!}U^{1,2}\left(\frac{\overleftarrow{\partial}}{\partial x}\frac{\overrightarrow{\partial}}{\partial \ddot{p}}\right)^n\left(\frac{\overleftarrow{\partial}}{\partial p}\frac{\overrightarrow{\partial}}{\partial \dot{p}}\right)^{2l-n+1}W^{1,2,3,4}. \quad (4.13)$$

Considering the square-law characteristic of potential $U^{1,2}$ (4.6), equation (4.13) for function (4.11) takes the form:

$$\left[\frac{\partial}{\partial t} + \frac{p}{m}\frac{\partial}{\partial x} + \dot{p}\frac{\partial}{\partial p} + \left(\ddot{p} - \frac{\partial}{\partial v}U^{1,2}\right)\frac{\partial}{\partial \dot{p}} + \frac{\partial}{\partial x}U^{1,2}\frac{\partial}{\partial \ddot{p}}\right]W^{1,2,3,4} = 0. \quad (4.14)$$

**Conclusions**

A new method has been proposed to obtain exact solutions of the Schrödinger equation on the base of its relation (i.2)-(i.14) with the Vlasov equations and with the Moyal equation. The solutions obtained can correspond to both classical and quantum models. The used apparatus of the Wigner function made it possible to visually interpret the time-dependent processes in a quantum system in terms of the classical analysis of the system dynamics in the phase space.

The presented results are applicable when considering open quantum systems, dissipative processes, and analysis of dynamical systems.

**Acknowledgements**

This research has been supported by the Interdisciplinary Scientific and Educational School of Moscow University «Photonic and Quantum Technologies. Digital Medicine».

**Appendix A**

*Proof of Theorem 1*

Equation (1.3) is a non-homogeneous first-order linear differential equation with variable coefficients. Let us calculate the partial derivatives included in the Vlasov equation:

$$\frac{\partial f_n^1}{\partial t} = -\dot{\sigma}\frac{1}{2^n n!}\frac{1}{\sqrt{2\pi\sigma^2}}\exp\left(-\frac{x^2}{2\sigma^2}\right)H_n^2\left(\frac{x}{\sqrt{2}\sigma}\right) + \frac{1}{2^n n!}\frac{1}{\sqrt{2\pi\sigma}}\frac{x^2}{\sigma^3}\dot{\sigma}\exp\left(-\frac{x^2}{2\sigma^2}\right)H_n^2\left(\frac{x}{\sqrt{2}\sigma}\right) +$$
$$-\frac{1}{2^n n!}\frac{1}{\sqrt{2\pi\sigma}}\exp\left(-\frac{x^2}{2\sigma^2}\right)H_n\left(\frac{x}{\sqrt{2}\sigma}\right)H_{n-1}\left(\frac{x}{\sqrt{2}\sigma}\right)2n\frac{2x\dot{\sigma}}{\sqrt{2}\sigma^2},$$

$$\frac{\partial f_n^1}{\partial t} = 2\left[\tilde{x}^2 - 2n\frac{H_{n-1}(\tilde{x})}{H_n(\tilde{x})}\tilde{x} - \frac{1}{2}\right]\frac{\dot{\sigma}}{\sigma}\tilde{f}_n^1, \quad (A.1)$$

$$\frac{\partial f_n^1}{\partial x} = \frac{2}{\sigma\sqrt{2}}\left[-\tilde{x} + 2n\frac{H_{n-1}(\tilde{x})}{H_n(\tilde{x})}\right]\tilde{f}_n^1, \quad (A.2)$$

where

$$f_n^1(x,t) = f_n^1\left(\tilde{x}\sigma(t)\sqrt{2},t\right) \stackrel{\text{det}}{=} \tilde{f}_n^1(\tilde{x},t), \qquad \tilde{x}(x,t) \stackrel{\text{det}}{=} \frac{x}{\sqrt{2}\sigma(t)}, \quad (A.3)$$

and it is taken into consideration that $H'_n(\tilde{x}) = 2nH_{n-1}(\tilde{x})$. Substituting expressions (A.1) and (A.2) into equation (1.2), we obtain equation (1.3).

The solution of equation (1.3) can be represented as:

$$\langle \tilde{v} \rangle_1 = \langle \tilde{v} \rangle_{g.h} + \langle \tilde{v} \rangle_{p.u}, \qquad (A.4)$$

where $\langle \tilde{v} \rangle_{g.h}$ as the general solution of the homogeneous equation:

$$\frac{1}{2}\frac{\partial \langle \tilde{v} \rangle_1}{\partial \tilde{x}} + \langle \tilde{v} \rangle_1 \left[ -\tilde{x} + 2n\frac{H_{n-1}(\tilde{x})}{H_n(\tilde{x})} \right] = 0, \qquad (A.5)$$

having the form:

$$\langle \tilde{v} \rangle_{g.h}(\tilde{x},t) = C \cdot \exp\left\{ -2\int \left[ \ln'|H_n(\tilde{x})| - \tilde{x} \right] d\tilde{x} \right\} = C \cdot \exp\left[ -2\ln|H_n(\tilde{x})| + \tilde{x}^2 \right],$$

$$\langle \tilde{v} \rangle_{g.h}(\tilde{x},t) = \frac{C}{H_n^2(\tilde{x})}\exp(\tilde{x}^2). \qquad (A.6)$$

A particular solution $\langle \tilde{v} \rangle_{p.u}$ of the inhomogeneous equation (1.3) can be represented as:

$$\langle \tilde{v} \rangle_{p.u}(\tilde{x},t) = \sqrt{2}\dot{\sigma}(t)\tilde{x}. \qquad (A.7)$$

Substituting (A.6) and (A.7) into representation (A.4) gives the general solution of equation (1.3). Theorem 1 is proved.

**Appendix B**

*Proof of Theorem 2*

Let us calculate the summands in expression (2.6):

$$\frac{\partial \varphi_n^1}{\partial t} = -\frac{x^2}{4\alpha}\frac{\ddot{\sigma}\sigma - \dot{\sigma}^2}{\sigma^2} - \beta \dot{E}_n, \quad \frac{\partial \varphi_n^1}{\partial x} = -\frac{x}{2\alpha}\frac{\dot{\sigma}}{\sigma}, \qquad (B.1)$$

$$\frac{\partial}{\partial \tilde{x}}\left[ e^{-\frac{\tilde{x}^2}{2}} H_n(\tilde{x}) \right] = -\tilde{x}e^{-\frac{\tilde{x}^2}{2}} H_n(\tilde{x}) + e^{-\frac{\tilde{x}^2}{2}} 2nH_{n-1}(\tilde{x}) = e^{-\frac{\tilde{x}^2}{2}}\left[ -\tilde{x}H_n(\tilde{x}) + 2nH_{n-1}(\tilde{x}) \right],$$

$$\frac{\partial^2}{\partial \tilde{x}^2}\left[ e^{-\frac{\tilde{x}^2}{2}} H_n(\tilde{x}) \right] = e^{-\frac{\tilde{x}^2}{2}}\left[ (\tilde{x}^2 - 1)H_n(\tilde{x}) - 4n\tilde{x}H_{n-1}(\tilde{x}) + 4n(n-1)H_{n-2}(\tilde{x}) \right]. \qquad (B.2)$$

Given the recurrence relation for the Hermitian polynomials in expression (B.2):

$$H_n(\tilde{x}) = 2\tilde{x}H_{n-1}(\tilde{x}) - 2(n-1)H_{n-2}(\tilde{x}), \; n \geq 2, \qquad (B.3)$$

we obtain the expression for quantum potential $Q^1$ (i.7):

$$\frac{1}{\sqrt{f_n^1}}\frac{\partial^2 \sqrt{f_n^1}}{\partial x^2} = \frac{1}{2\sigma^2} = \frac{1}{2\sigma^2}\frac{e^{\frac{\tilde{x}^2}{2}}}{H_n(\tilde{x})}\frac{\partial^2}{\partial \tilde{x}^2}\left[e^{-\frac{\tilde{x}^2}{2}}H_n(\tilde{x})\right] =$$

$$= \frac{1}{2\sigma^2}\frac{1}{H_n(\tilde{x})}\left\{(\tilde{x}^2-1)H_n(\tilde{x}) - 4n\left[\tilde{x}H_{n-1}(\tilde{x}) - (n-1)H_{n-2}(\tilde{x})\right]\right\} =$$

$$= \frac{1}{2\sigma^2}\frac{1}{H_n(\tilde{x})}\left[(\tilde{x}^2-1)H_n(\tilde{x}) - 2nH_n(\tilde{x})\right],$$

$$\frac{1}{\sqrt{f_n^1}}\frac{\partial^2 \sqrt{f_n^1}}{\partial x^2} = \frac{\tilde{x}^2 - 1 - 2n}{2\sigma^2}. \tag{B.4}$$

By substituting (B.1) and (B.4) into expression (2.6), we obtain:

$$U^1(x,t) = \frac{x^2}{4\alpha\beta}\frac{\ddot{\sigma}\sigma - \dot{\sigma}^2}{\sigma^2} + \dot{E}_n - \frac{\alpha}{\beta}\frac{\tilde{x}^2 - 1 - 2n}{2\sigma^2} + \frac{x^2}{4\alpha\beta}\frac{\dot{\sigma}^2}{\sigma^2} = \frac{\ddot{\sigma}x^2}{4\alpha\beta\sigma} - \frac{\alpha}{\beta}\frac{\tilde{x}^2}{2\sigma^2} + \dot{E}_n + \frac{\alpha}{\beta}\frac{1+2n}{2\sigma^2},$$

$$U^1(x,t) = \frac{1}{4\alpha\beta\sigma}\left(\ddot{\sigma} - \frac{\alpha^2}{\sigma^3}\right)x^2 + \dot{E}_n + \frac{\alpha}{\beta}\frac{1+2n}{2\sigma^2}. \tag{B.5}$$

Theorem 2 is proved.

We obtain the equation for the separatrix from condition $\Omega(t) = 0$:

$$\ddot{\sigma}\sigma = \frac{\alpha^2}{\sigma^3}\dot{\sigma} \Rightarrow \frac{d}{dt}\dot{\sigma}^2 = -\alpha^2\frac{d}{dt}\frac{1}{\sigma^2} \Rightarrow \dot{\sigma}^2 = c_1^2 - \frac{\alpha^2}{\sigma^2} \Rightarrow \frac{\sigma d\sigma}{\sqrt{c_1^2\sigma^2 - \alpha^2}} = \pm dt,$$

$$\pm t + c_2 = \int \frac{\sigma d\sigma}{\sqrt{c_1^2\sigma^2 - \alpha^2}} = \frac{1}{2c_1^2}\int \frac{d(c_1^2\sigma^2 - \alpha^2)}{\sqrt{c_1^2\sigma^2 - \alpha^2}} = \frac{1}{c_1^2}\sqrt{c_1^2\sigma^2 - \alpha^2},$$

$$\pm c_1^2 t + c_2 = \sqrt{c_1^2\sigma^2 - \alpha^2} \Rightarrow (\pm c_1^2 t + c_2)^2 + \alpha^2 = c_1^2\sigma^2 \Rightarrow \sigma^2(t) = (\pm c_1 t + c_2)^2 + \frac{\alpha^2}{c_1^2},$$

$$\sigma^2(t) = c_1^2(t \pm c_2)^2 + \frac{\alpha^2}{c_1^2}. \tag{B.6}$$

**Appendix C**

*Proof of Theorem 3*

Let us substitute the expression for wave function (2.11) into integral (3.1). By transforming the expression under integral sign (3.1), we obtain:

$$2^n n!\sqrt{2\pi}\sigma \,\overline{\Psi}_n^1(x',t)\Psi_n^1(x'',t)\exp\left(-i\frac{ps}{\hbar}\right) =$$

$$= \exp\left[-\frac{1}{4\sigma^2}(x'^2 + x''^2) + i\frac{\dot{\sigma}}{4\alpha\sigma}(x'^2 - x''^2) - i\frac{ps}{\hbar}\right]H_n\left(\frac{x'}{\sqrt{2}\sigma}\right)H_n\left(\frac{x''}{\sqrt{2}\sigma}\right) = \tag{C.1}$$

$$= \exp\left[-\frac{1}{2\sigma^2}\left(x^2 + \frac{s^2}{4}\right) - i\frac{\dot{\sigma}}{2\alpha\sigma}xs - i\frac{ps}{\hbar}\right]H_n\left(\frac{x'}{\sqrt{2}\sigma}\right)H_n\left(\frac{x''}{\sqrt{2}\sigma}\right),$$

where $x' = x - \dfrac{s}{2}$, $x'' = x + \dfrac{s}{2}$ and it is taken into account that $x'^2 + x''^2 = 2x^2 + \dfrac{s^2}{2}$ and $x'^2 - x''^2 = -2xs$. Let us substitute expression (C.1) into integral (3.1)

$$W_n(x,p,t) = \frac{e^{-\kappa^2 x^2}}{2^{n+1} n! \pi^{3/2} \hbar} \int_{-\infty}^{+\infty} e^{-\frac{\xi^2}{4} - i\left(\frac{\sigma\dot\sigma}{\alpha}\kappa x + \frac{p}{\kappa\hbar}\right)\xi} H_n\left(\kappa x - \frac{\xi}{2}\right) H_n\left(\kappa x + \frac{\xi}{2}\right) d\xi. \qquad (C.2)$$

where $\kappa = \dfrac{1}{\sigma\sqrt{2}}$, $\xi = \kappa s$. We transform the integral in expression (C.2):

$$\int_{-\infty}^{+\infty} e^{-\frac{\xi^2}{4} - i\left(\frac{\sigma\dot\sigma}{\alpha}\kappa x + \frac{p}{\kappa\hbar}\right)\xi} H_n\left(\kappa x - \frac{\xi}{2}\right) H_n\left(\kappa x + \frac{\xi}{2}\right) d\xi =$$

$$= 2e^{-\left(\frac{\sigma\dot\sigma}{\alpha}\kappa x + \frac{p}{\kappa\hbar}\right)^2} \int_{-\infty}^{+\infty} e^{-\varsigma^2} H_n\left(\kappa x + i\left(\frac{\sigma\dot\sigma}{\alpha}\kappa x + \frac{p}{\kappa\hbar}\right) - \varsigma\right) H_n\left(\kappa x - i\left(\frac{\sigma\dot\sigma}{\alpha}\kappa x + \frac{p}{\kappa\hbar}\right) + \varsigma\right) d\varsigma = \quad (C.3)$$

$$= 2(-1)^n e^{-\left(\frac{\sigma\dot\sigma}{\alpha}\kappa x + \frac{p}{\kappa\hbar}\right)^2} \int_{-\infty}^{+\infty} e^{-\varsigma^2} H_n(\varsigma + \varsigma_1) H_n(\varsigma + \varsigma_2) d\varsigma,$$

where $\varsigma \overset{\text{det}}{=} \dfrac{\xi}{2} + i\left(\dfrac{\sigma\dot\sigma}{\alpha}\kappa x + \dfrac{p}{\kappa\hbar}\right)$, $\varsigma_1 \overset{\text{det}}{=} \kappa x - i\left(\dfrac{\sigma\dot\sigma}{\alpha}\kappa x + \dfrac{p}{\kappa\hbar}\right)$, $\varsigma_2 \overset{\text{det}}{=} -\kappa x - i\left(\dfrac{\sigma\dot\sigma}{\alpha}\kappa x + \dfrac{p}{\kappa\hbar}\right)$ and the property of Hermitian polynomials $H_n(-x) = (-1)^n H_n(x)$ is taken into account, i.e.,

$$H_n\left(-\left(\varsigma - \kappa x - i\left(\frac{\sigma\dot\sigma}{\alpha}\kappa x + \frac{p}{\kappa\hbar}\right)\right)\right) = (-1)^n H_n\left(\varsigma - \kappa x - i\left(\frac{\sigma\dot\sigma}{\alpha}\kappa x + \frac{p}{\kappa\hbar}\right)\right).$$

Integral (C.3) can be expressed in terms of the Laguerre polynomials [33]

$$\frac{1}{2^n n! \sqrt{\pi}} \int_{-\infty}^{+\infty} e^{-\varsigma^2} H_n(\varsigma + \varsigma_1) H_n(\varsigma + \varsigma_2) d\varsigma = L_n(-2\varsigma_1\varsigma_2). \qquad (C.4)$$

Given representation (C.4) in expression (C.3), we obtain the Wigner function (C.2)

$$W_n(x,p,t) = \frac{e^{-\kappa^2 x^2} 2^n n! \sqrt{\pi}}{2^{n+1} n! \pi^{3/2} \hbar} 2(-1)^n e^{-\left(\frac{\sigma\dot\sigma}{\alpha}\kappa x + \frac{p}{\kappa\hbar}\right)^2} L_n(-2\varsigma_1\varsigma_2) =$$

$$= \frac{(-1)^n}{\pi\hbar} e^{-\kappa^2 x^2 - \left(\frac{\sigma\dot\sigma}{\alpha}\kappa x + \frac{p}{\kappa\hbar}\right)^2} L_n(-2\varsigma_1\varsigma_2). \qquad (C.5)$$

Note that $-\varsigma_2 = \overline{\varsigma_1}$, therefore, the argument of the Laguerre polynomials $-2\varsigma_1\varsigma_2 = 2\varsigma_1\overline{\varsigma_1} = 2|\varsigma_1|^2$ is, i.e,

$$W_n(x,p,t) = \frac{(-1)^n}{\pi\hbar} e^{-\kappa^2 x^2 - \left(\frac{\sigma\dot\sigma}{\alpha}\kappa x + \frac{p}{\kappa\hbar}\right)^2} L_n\left(2\kappa^2 x^2 + 2\left(\frac{\sigma\dot\sigma}{\alpha}\kappa x + \frac{p}{\kappa\hbar}\right)^2\right). \qquad (C.6)$$

Theorem 3 is proved.

We write the second-order curve equation (3.5) in dimensionless coordinates $\tilde{x} \stackrel{\text{det}}{=} \kappa x$ and $\tilde{p} \stackrel{\text{det}}{=} \dfrac{p}{\hbar\kappa}$, we obtain

$$a_{11}\tilde{x}^2 + 2a_{12}\tilde{x}\tilde{p} + a_{22}\tilde{p}^2 = const, \tag{C.7}$$

where

$$a_{11} = 1 + \frac{\sigma^2\dot{\sigma}^2}{\alpha^2}, \quad a_{12} = a_{21} = \frac{\sigma\dot{\sigma}}{\alpha}, \quad a_{22} = 1.$$

Therefore, angle of inclination $\vartheta$ of the principal axes of the ellipse (C.7) satisfies the condition:

$$\text{tg}(2\vartheta) = \frac{2a_{12}}{a_{11}-a_{22}} = \frac{2\alpha}{\sigma\dot{\sigma}} = -\frac{\hbar}{m\sigma\dot{\sigma}}, \tag{C.8}$$

and its area (at $const = 1$)

$$S = \frac{\pi}{\sqrt{a_{11}a_{22} - a_{12}^2}} = \pi.$$

Let us verify that function $\varepsilon(x,p,t)$ is a characteristic of the Moyal equation (3.9). We make direct calculations:

$$\left[\frac{\partial}{\partial t} + \frac{p}{m}\frac{\partial}{\partial x} - \frac{\partial}{\partial x}U^1\frac{\partial}{\partial p}\right]W_n = \frac{(-1)^n}{\pi\hbar}\left[-L_n(2\varepsilon) + 2L_n'(2\varepsilon)\right]e^{-\varepsilon}\left[\frac{\partial\varepsilon}{\partial t} + \frac{p}{m}\frac{\partial\varepsilon}{\partial x} - m\Omega^2 x\frac{\partial\varepsilon}{\partial p}\right] = 0. \tag{C.9}$$

The partial derivatives have the form:

$$\frac{\partial\varepsilon}{\partial t} = \dot{\sigma}\left(\frac{\ddot{\sigma}}{\alpha^2} - \frac{1}{\sigma^3}\right)x^2 + \frac{4\sigma\dot{\sigma}}{\hbar^2}p^2 + \frac{2}{\alpha\hbar}(\dot{\sigma}^2 + \sigma\ddot{\sigma})xp,$$

$$\frac{\partial\varepsilon}{\partial x} = \frac{x}{\sigma} + 2\frac{\dot{\sigma}}{\alpha\sqrt{2}}\left(\frac{\dot{\sigma}}{\alpha\sqrt{2}}x + \frac{\sigma\sqrt{2}}{\hbar}p\right) = \left(\frac{1}{\sigma} + \frac{\dot{\sigma}^2}{\alpha^2}\right)x + \frac{2\sigma\dot{\sigma}}{\alpha\hbar}p, \tag{C.10}$$

$$\frac{\partial\varepsilon}{\partial p} = 2\left(\frac{\dot{\sigma}}{\alpha\sqrt{2}}x + \frac{\sigma\sqrt{2}}{\hbar}p\right)\frac{\sigma\sqrt{2}}{\hbar} = \frac{2\sigma\dot{\sigma}}{\alpha\hbar}x + \frac{4\sigma^2}{\hbar^2}p.$$

Substituting expressions (C.10) into equation (C.9), we obtain

$$\frac{\partial\varepsilon}{\partial t} + \frac{p}{m}\frac{\partial\varepsilon}{\partial x} - m\Omega^2 x\frac{\partial\varepsilon}{\partial p} = \dot{\sigma}\left(\frac{4m^2\ddot{\sigma}}{\hbar^2} - \frac{1}{\sigma^3}\right)x^2 + \frac{4\sigma\dot{\sigma}}{\hbar^2}p^2 - \frac{4m}{\hbar^2}(\dot{\sigma}^2 + \sigma\ddot{\sigma})xp +$$

$$+ \frac{p}{m}\left(\frac{1}{\sigma} + \frac{4m^2\dot{\sigma}^2}{\hbar^2}\right)x - \frac{p}{m}\frac{4m\sigma\dot{\sigma}}{\hbar^2}p - mx\frac{1}{\sigma}\left(\frac{\alpha^2}{\sigma^3} - \ddot{\sigma}\right)\left(-\frac{4m\sigma\dot{\sigma}}{\hbar^2}x + \frac{4\sigma^2}{\hbar^2}p\right) =$$

$$= \frac{4\sigma\dot{\sigma}}{\hbar^2}p^2 - \frac{4\sigma\dot{\sigma}}{\hbar^2}p^2 + \left[\frac{4m^2\ddot{\sigma}\dddot{\sigma}}{\hbar^2} - \frac{\dot{\sigma}}{\sigma^3} + \frac{\dot{\sigma}}{\sigma^3} - \frac{4m^2\ddot{\sigma}\dddot{\sigma}}{\hbar^2}\right]x^2 +$$

$$+\left[\frac{1}{m\sigma^2} - \frac{1}{m\sigma^2} + \frac{4m\dot{\sigma}^2}{\hbar^2} - \frac{4m\dot{\sigma}^2}{\hbar^2} + \frac{4m\sigma\ddot{\sigma}}{\hbar^2} - \frac{4m\sigma\ddot{\sigma}}{\hbar^2}\right]xp = 0,$$

which was to be proved.

## Appendix D

### *Proof of Theorem 4*

Potential $U^{1,2}(x,v,t)$ can be found in two ways. The first way is to express from the Hamilton-Jacobi equation (4.7)-(4.8), and the second – from the Schrödinger equation (4.5). In both cases, the same result will be obtained. Without loss of generality, let us use the second method. We calculate the derivatives:

$$\frac{1}{\Psi^{1,2}}\frac{\partial \Psi^{1,2}}{\partial t} = \frac{\dot{\sigma}x^2}{2\sigma^3} - 2\frac{m^2}{\hbar^2}(\dot{\sigma}x - \sigma v)(\ddot{\sigma}x - \dot{\sigma}v) + i\left(\frac{\hbar^2}{m\sigma^5}\dot{\sigma} + m\frac{\ddot{\sigma}\sigma - \dot{\sigma}\dot{\sigma}}{\sigma^2}\right)\frac{xv}{\hbar_2} - i\frac{\dot{E}^{1,2}}{\hbar_2}, \qquad (D.1)$$

$$\frac{1}{\Psi^{1,2}}\frac{\partial \Psi^{1,2}}{\partial x} = -\frac{x}{2\sigma^2} - 2\frac{m^2}{\hbar^2}(\dot{\sigma}x - \sigma v)\dot{\sigma} - i\left(\frac{\hbar^2}{2m\sigma^4} - 2m\frac{\ddot{\sigma}}{\sigma}\right)\frac{v}{2\hbar_2}, \qquad (D.2)$$

$$\frac{1}{\Psi^{1,2}}\frac{\partial \Psi^{1,2}}{\partial v} = 2\frac{m^2}{\hbar^2}(\dot{\sigma}x - \sigma v)\sigma - i\left(\frac{\hbar^2}{2m\sigma^4} - 2m\frac{\ddot{\sigma}}{\sigma}\right)\frac{x}{2\hbar_2},$$

$$\frac{1}{\Psi^{1,2}}\frac{\partial^2 \Psi^{1,2}}{\partial v^2} = \left[-2\frac{m^2}{\hbar^2}(\dot{\sigma}x - \sigma v)\sigma + i\left(\frac{\hbar^2}{2m\sigma^4} - 2m\frac{\ddot{\sigma}}{\sigma}\right)\frac{x}{2\hbar_2}\right]^2 - 2\frac{m^2\sigma^2}{\hbar^2} = -2\frac{m^2\sigma^2}{\hbar^2} + \qquad (D.3)$$

$$+4\frac{m^4}{\hbar^4}(\sigma v - \dot{\sigma}x)^2\sigma^2 - \left(\frac{\hbar^2}{2m\sigma^4} - 2m\frac{\ddot{\sigma}}{\sigma}\right)^2\frac{x^2}{4\hbar_2^2} + 2i\frac{m^2}{\hbar^2\hbar_2}(\sigma v - \dot{\sigma}x)\sigma\left(\frac{\hbar^2}{2m\sigma^4} - 2m\frac{\ddot{\sigma}}{\sigma}\right)x.$$

Taking into account (D.1) and (D.2), we find derivative $\partial_1 \Psi^{1,2}$:

$$-\frac{1}{\Psi^{1,2}}\left(\frac{\partial}{\partial t} + v\frac{\partial}{\partial x}\right)\Psi^{1,2} = \left(2\frac{m^2}{\hbar^2}\dot{\sigma}\ddot{\sigma} - \frac{\dot{\sigma}}{2\sigma^3}\right)x^2 + \left(\frac{1}{2\sigma^2} - 2\frac{m^2}{\hbar^2}\ddot{\sigma}\sigma\right)xv +$$

$$+ i\left(\frac{\hbar^2}{2m\sigma^4} - 2m\frac{\ddot{\sigma}}{\sigma}\right)\frac{v^2}{2\hbar_2} - i\left(\frac{\hbar^2\dot{\sigma}}{m\sigma^5} + m\frac{\ddot{\sigma}\sigma - \dot{\sigma}\dot{\sigma}}{\sigma^2}\right)\frac{xv}{\hbar_2} + i\frac{\dot{E}^{1,2}}{\hbar_2}. \qquad (D.4)$$

Substituting (D.3) and (D.4) into equation (4.5), we obtain the expression for potential $U^{1,2}(x,v,t)$:

$$U^{1,2} = \frac{i\hbar_2}{\Psi^{1,2}}\left(\frac{\partial}{\partial t}+v\frac{\partial}{\partial x}\right)\Psi^{1,2} + \frac{1}{\Psi^{1,2}}\frac{\hbar_2^2}{2m}\frac{\partial^2}{\partial v^2}\Psi^{1,2} = \left(\frac{\hbar^2}{2m\sigma^4}-2m\frac{\ddot{\sigma}}{\sigma}\right)\frac{v^2}{2} -$$

$$-\left(\frac{\hbar^2\dot{\sigma}}{m\sigma^5}+m\frac{\dddot{\sigma}\sigma-\ddot{\sigma}\dot{\sigma}}{\sigma^2}\right)xv - \frac{m\hbar_2^2\sigma^2}{\hbar^2} + 2\frac{\hbar_2^2 m^3}{\hbar^4}(\sigma v-\dot{\sigma}x)^2\sigma^2 - \left(\frac{\hbar^2}{2m\sigma^4}-2m\frac{\ddot{\sigma}}{\sigma}\right)^2\frac{x^2}{8m} + \dot{E}^{1,2} -$$

$$-i\hbar_2\left[\left(2\frac{m^2}{\hbar^2}\ddot{\sigma}-\frac{1}{2\sigma^3}+\frac{1}{2\sigma^3}-2\frac{m^2}{\hbar^2}\ddot{\sigma}\right)\dot{\sigma}x^2 + \left(\frac{1}{2\sigma^2}-2\frac{m^2}{\hbar^2}\ddot{\sigma}\sigma\right)xv - \left(\frac{1}{2\sigma^2}-2\frac{m^2}{\hbar^2}\sigma\ddot{\sigma}\right)xv\right] =$$

$$U^{1,2} = \left(\frac{\hbar^2}{4m\sigma^4}-m\frac{\ddot{\sigma}}{\sigma}+2\frac{\hbar_2^2 m^3\sigma^4}{\hbar^4}\right)v^2 - \left(\frac{\hbar^2\dot{\sigma}}{m\sigma^5}+m\frac{\dddot{\sigma}\sigma-\ddot{\sigma}\dot{\sigma}}{\sigma^2}+4\frac{\hbar_2^2 m^3}{\hbar^4}\sigma^3\dot{\sigma}\right)xv +$$

(D.5)

$$+\left[2\frac{\hbar_2^2 m^3}{\hbar^4}\sigma^2\dot{\sigma}^2 - \left(\frac{\hbar^2}{4m\sqrt{2m}\sigma^4}-\frac{\ddot{\sigma}}{\sigma}\sqrt{\frac{m}{2}}\right)^2\right]x^2 + \dot{E}^{1,2} - \frac{m\hbar_2^2\sigma^2}{\hbar^2}.$$

Since function $E^{1,2}$ is arbitrary, then we set the condition for it $\dot{E}^{1,2}(t) = \frac{m\hbar_2^2}{\hbar^2}\sigma^2(t)$, then expression (D.5) will go into the required one (4.6). Theorem 4 is proved.

*Proof of Theorem 5*

Let us find the Wigner function of the fourth rank. Let us transform the expression under integral sign in the Fourier transform (4.10), we obtain

$$\pi\hbar\overline{\Psi}^{1,2}(x',v',t)\Psi^{1,2}(x'',v'',t) = \qquad (D.6)$$

$$= \exp\left[-\frac{1}{4\sigma^2}(x'^2+x''^2) - \frac{m^2}{\hbar^2}\left[(\dot{\sigma}x'-\sigma v')^2+(\dot{\sigma}x''-\sigma v'')^2\right] + \frac{i}{2\hbar_2}\left(\frac{\hbar^2}{2m\sigma^4}-2m\frac{\ddot{\sigma}}{\sigma}\right)(x'v'-x''v'')\right]$$

Let us make intermediate transformations:

$$x'^2+x''^2 = 2x^2+\frac{s_1^2}{2}, \quad x'v'-x''v'' = -xs_2-s_1v, \qquad (D.7)$$

$$(\dot{\sigma}x'-\sigma v')^2+(\dot{\sigma}x''-\sigma v'')^2 = 2(\dot{\sigma}x-\sigma v)^2+\frac{1}{2}(\dot{\sigma}s_1-\sigma s_2)^2.$$

Considering (D.7), expression (D.6) will take the form:

$$\pi\hbar\overline{\Psi}^{1,2}(x',v',t)\Psi^{1,2}(x'',v'',t)\exp\left(i\frac{s_1\ddot{p}-s_2\dot{p}}{\hbar_2}\right) = \exp\left[-\frac{x^2}{2\sigma^2}-2\frac{m^2}{\hbar^2}(\dot{\sigma}x-\sigma v)^2\right]\times \qquad (D.8)$$

$$\times\exp\left\{-\frac{s_1^2}{8\sigma^2}-\frac{m^2}{2\hbar^2}(\dot{\sigma}s_1-\sigma s_2)^2 - \frac{i}{\hbar_2}\left[\left(\frac{\hbar^2}{4m\sigma^4}-m\frac{\ddot{\sigma}}{\sigma}\right)x+\dot{p}\right]s_2 - \frac{i}{\hbar_2}\left[\left(\frac{\hbar^2}{4m\sigma^4}-m\frac{\ddot{\sigma}}{\sigma}\right)v-\ddot{p}\right]s_1\right\}$$

The first exponent in expression (D.8) does not participate in integration (4.10), so we transform the second exponent into (D.8)

$$-\frac{s_1^2}{8\sigma^2} - \frac{m^2}{2\hbar^2}(\dot{\sigma}s_1 - \sigma s_2)^2 - \frac{i}{\hbar_2}\left[\left(\frac{\hbar^2}{4m\sigma^4} - m\frac{\ddot{\sigma}}{\sigma}\right)x + \dot{p}\right]s_2 - \frac{i}{\hbar_2}\left[\left(\frac{\hbar^2}{4m\sigma^4} - m\frac{\ddot{\sigma}}{\sigma}\right)v - \ddot{p}\right]s_1 =$$
$$= -A(t)s_1^2 + B(s_2, v, \ddot{p}, t)s_1 - c_1(t)s_2^2 - ic_2(x, \dot{p}, t)s_2, \qquad (D.9)$$

where

$$A(t) \stackrel{\text{det}}{=} \frac{1}{8\sigma^2} + \frac{m^2\dot{\sigma}^2}{2\hbar^2}, \quad c_1(t) \stackrel{\text{det}}{=} \frac{m^2\sigma^2}{2\hbar^2}, \quad c_2(x, \dot{p}, t) \stackrel{\text{det}}{=} \frac{1}{\hbar_2}\left[\left(\frac{\hbar^2}{4m\sigma^4} - m\frac{\ddot{\sigma}}{\sigma}\right)x + \dot{p}\right], \qquad (D.10)$$

$$B(s_2, v, \ddot{p}, t) \stackrel{\text{det}}{=} b_1(t)s_2 + ib_2, \quad b_1(t) \stackrel{\text{det}}{=} \frac{m^2\sigma\dot{\sigma}}{\hbar^2}, \quad b_2(v, \ddot{p}, t) \stackrel{\text{det}}{=} \frac{1}{\hbar_2}\left[\ddot{p} - \left(\frac{\hbar^2}{4m\sigma^4} - m\frac{\ddot{\sigma}}{\sigma}\right)v\right].$$

We select the perfect square in expression (D.9), we obtain:

$$-As_1^2 + Bs_1 - c_1 s_2^2 - ic_2 s_2 = -A\left(s_1 - \frac{B}{2A}\right)^2 + \frac{b_1^2 - 4Ac_1}{4A}\left(s_2 + i\frac{b_1 b_2 - 2Ac_2}{b_1^2 - 4Ac_1}\right)^2 +$$
$$+ \frac{Ac_2^2 - b_1 b_2 c_2 + b_2^2 c_1}{b_1^2 - 4Ac_1}. \qquad (D.11)$$

The first two summands in expression (D.11) participate in the integration (4.10), and the third one is taken out of the integral sign. We transform the third summand in expression (D.11) using the notation from (D.10)

$$-\frac{m^2 \hbar_2^2}{4} \frac{Ac_2^2 - b_1 b_2 c_2 + b_2^2 c_1}{b_1^2 - 4Ac_1} = \frac{\hbar^2 + 4m^2\sigma^2\dot{\sigma}^2}{8\sigma^2}\left(\dot{p} + \frac{\hbar^2 - 4m^2\sigma^3\ddot{\sigma}}{4m\sigma^4}x\right)^2 +$$
$$+ \left[\frac{\sigma}{2}\left(\ddot{p} - \frac{\hbar^2 - 4m^2\sigma^3\ddot{\sigma}}{4m\sigma^4}v\right) - \dot{\sigma}\left(\dot{p} + \frac{\hbar^2 - 4m^2\sigma^3\ddot{\sigma}}{4m\sigma^4}x\right)\right]\left(\ddot{p} - \frac{\hbar^2 - 4m^2\sigma^3\ddot{\sigma}}{4m\sigma^4}v\right)m^2\sigma. \qquad (D.12)$$

Let us introduce the notation: $\xi(t) \stackrel{\text{det}}{=} \frac{\hbar^2 + 4m^2\sigma^2\dot{\sigma}^2}{4m^2\sigma^2\dot{\sigma}^2}$ and $\eta(t) \stackrel{\text{det}}{=} \frac{\hbar^2 - 4m^2\sigma^3\ddot{\sigma}}{4m\sigma^4}$, then expression (D.12) will take the form:

$$\frac{Ac_2^2 - b_1 b_2 c_2 + b_2^2 c_1}{b_1^2 - 4Ac_1} = -\frac{2}{m^2 \hbar_2^2}\left(\xi\dot{P}^2 - 2\ddot{P}\dot{P} + \ddot{P}^2\right), \qquad (D.13)$$

where

$$\dot{P}(x, \dot{p}, t) \stackrel{\text{det}}{=} m\dot{\sigma}(t)\left[\dot{p} + \eta(t)x\right], \quad \ddot{P}(v, \ddot{p}, t) \stackrel{\text{det}}{=} m\sigma(t)\left[\ddot{p} - \eta(t)v\right]. \qquad (D.14)$$

Substituting expression (D.13) into (D.11), and then into (D.8) and finally into integral (4.10), we obtain

$$W^{1,2,3,4}(x,p,\dot{p},\ddot{p}) = \frac{m}{\pi\hbar(2\pi\hbar_2)^2} e^{-\frac{x^2}{2\sigma^2} - 2\frac{m^2}{\hbar^2}(\dot\sigma x - \sigma v)^2} e^{-\frac{2}{m^2\hbar_2^2}(\xi\dot P^2 - 2\ddot P\dot P + \ddot P^2)} \times$$

$$\times \int_{-\infty}^{+\infty} e^{-A\left(s_1 - \frac{B}{2A}\right)^2} ds_1 \int_{-\infty}^{+\infty} e^{\frac{b_1^2 - 4Ac_1}{4A}\left(s_2 + i\frac{b_1 b_2 - 2Ac_2}{b_1^2 - 4Ac_1}\right)^2} ds_2,$$

$$W^{1,2,3,4}(x,p,\dot{p},\ddot{p}) = \frac{1}{(\pi\hbar_2)^2} \exp\left\{-\frac{x^2}{2\sigma^2} - 2\frac{m^2}{\hbar^2}(\dot\sigma x - \sigma v)^2 - \frac{4}{m^2\hbar_2^2}(\xi\dot P^2 - 2\ddot P\dot P + \ddot P^2)\right\}, \quad (D.15)$$

where it is taken into account that

$$\int_{-\infty}^{+\infty} e^{-A\left(s_1 - \frac{B}{2A}\right)^2} ds_1 = \sqrt{\frac{\pi}{A}} = 2\hbar\sigma\sqrt{\frac{2\pi}{\hbar^2 + 4m^2\sigma^2\dot\sigma^2}},$$

$$\int_{-\infty}^{+\infty} e^{\frac{b_1^2 - 4Ac_1}{4A}\left(s_2 + i\frac{b_1 b_2 - 2Ac_2}{b_1^2 - 4Ac_1}\right)^2} ds_2 = \frac{1}{m\sigma}\sqrt{2\pi(\hbar^2 + 4m^2\sigma^2\dot\sigma^2)}.$$

Expression (D.15) can be finally rewritten in terms of the original variables using the notation from (D.14)

$$\xi\dot P^2 - 2\ddot P\dot P + \ddot P^2 = \dot P^2 - 2\ddot P\dot P + \ddot P^2 + \frac{\hbar^2}{4m^2\sigma^2\dot\sigma^2}\dot P^2 = (\dot P - \ddot P)^2 + \frac{\hbar^2}{4m^2\sigma^2\dot\sigma^2}\dot P^2 =$$
$$= [m(\sigma\ddot p - \dot\sigma\dot p) - \eta(\sigma p + \dot\sigma mx)]^2 + \frac{\hbar^2}{4\sigma^2}(\dot p + \eta x)^2, \quad (D.16)$$

or

$$\xi\dot P^2 - 2\ddot P\dot P + \ddot P^2 = [m(\sigma\ddot p - \dot\sigma\dot p) - \eta(\sigma p + \dot\sigma mx)]^2 - \frac{\hbar^2}{4\sigma^2}(\dot p + \eta x)^2.$$

Substituting (D.16) into (D.15) we obtain expression (4.11). Theorem 5 is proved.

**References**

1. Daniel F. Styer and Miranda S. Balkin, Kathryn M. Becker, Matthew R. Burns, Christopher E. Dudley, Scott T. Forth, Jeremy S. Gaumer, Mark A. Kramer, David C. Oertel, Leonard H. Park, Marie T. Rinkoski, Clait T. Smith, and Timothy D. Wotherspoon, Nine formulations of quantum mechanics, Am. J. Phys. 70 (3), March 2002
2. L.D. Landau, E.M. Lifshitz, Quantum Mechanics, vol.3, Pergamon Press, 1977, 677 p.
3. T.C. Scott and Wenxing Zhang, Efficient hybrid-symbolic methods for quantum mechanical calculations, Comput. Phys. Commun., 2015, vol. 191, pp. 221-234.
4. G. V. Berghe, V. Fack, H.E. de Meyer, Numerical methods for solving radial Schrodinger equations, Journal of Computational and Applied Mathematics, 1989, vol. 28, pp. 391-401
5. John R. Ray, Exact solutions to the time-dependent Schrödinger equation, Phys. Rev. A, 1982, 26, 729
6. Millard F. Manning, Exact Solutions of the Schrödinger Equation, 1935, Phys. Rev. 48, 161
7. Guo-Hua Sun, Chang-Yuan Chen, Hind Taud, C. Yáñez-Márquez, Shi-Hai Dong, Exact solutions of the 1D Schrödinger equation with the Mathieu potential, Physics Letters A, 2020, vol. 384, Issue 19, № 126480
8. Simpao, Valentino A. (2014). «Real wave function from Generalised Hamiltonian Schrodinger Equation in quantum phase space via HOA (Heaviside Operational Ansatz):

exact analytical results». Journal of Mathematical Chemistry. 52 (4): 1137–1155. doi:10.1007/s10910-014-0332-2. ISSN 0259-9791.
9. R. C. Walker, A. W. Götz, Electronic Structure Calculations on Graphics Processing Units: From Quantum Chemistry to Condensed Matter Physics, 2016, ISBN:9781118661789, DOI:10.1002/9781118670712, John Wiley & Sons, Ltd
10. T.J. Barth, M. Gribel, D.E. Keyes, R.M. Nieminen, D. Roose, T. Schlick, From the Schrödinger Equation to Molecular Dynamics. In: Numerical Simulation in Molecular Dynamics. Texts in Computational Science and Engineering, (2007), vol 5. Springer, Berlin, Heidelberg. https://doi.org/10.1007/978-3-540-68095-6_2
11. M. Karplus, M. Levitt, A. Warshel, The Nobel Prize in Chemistry 2013.
12. A.A. Vlasov, Statistical Distribution Functions. Nauka, Moscow, 1966
13. Vlasov A.A., Many-Particle Theory and Its Application to Plasma, New York, Gordon and Breach, 1961, ISBN 0-677-20330-6
14. Perepelkin E.E., Sadovnikov B.I., Inozemtseva N.G. The properties of the first equation of the Vlasov chain of equations //J. Stat. Mech. – 2015. – №. P05019.
15. Perepelkin E.E., Sadovnikov B.I., Inozemtseva N.G., Korepanova A.A., Dispersion chain of quantum mechanics equations, 2023, Journal of Physics A: Mathematical and Theoretical , 2023, v. 56, №14, p. 1455202-1–145202-42.
16. Bohm, D., Hiley, B.J., Kaloyerou, P.N. (1987). An ontological basis for the quantum theory. *Phys. Rep.* 144: 321–375.
17. Bohm, D., Hiley, B.J. (1993). The Undivided Universe: An Ontological Interpretation of Quantum Theory. *Routledge, London*.
18. de Broglie, L. (1956). Une interpretation causale et non lineaire de la mecanique ondulatoire: la theorie de ladouble solution, *Gauthiers-Villiars, Paris*. (1956). English translation: Elsevier, Amsterdam (1960).
19. Perepelkin E.E., Sadovnikov B.I., Inozemtseva N.G., Burlakov E.V., Wigner function of a quantum system with polynomial potential //Journal of Statistical Mechanics: Theory and Experiment. – 2020. – №. 053105.
20. Moyal E. Quantum mechanics as a statistical theory //Proceedings of the Cambridge Philosophical Society. – 1949. – vol. 45. – pp. 99-124.
21. Wigner E.P. On the quantum correction for thermodynamic equilibrium, Phys. Rev., 1932. vol. 40, pp. 749-759.
22. H. Weyl, The theory of groups and quantum mechanics (Dover, New York, 1931).
23. Perepelkin E.E., Sadovnikov B.I., Inozemtseva N.G., Aleksandrov I.I., Dispersion chain of Vlasov equations //J. Stat. Mech. – 2022. – №. 0132
24. Perepelkin E.E., Sadovnikov B.I., Inozemtseva N.G., Ψ-model of micro- and macrosystems, Annals of Physics, 383 (2017) 511-544
25. Perepelkin E.E., Sadovnikov B.I., Inozemtseva N.G., Aleksandrov I.I., Exact time-dependent solution of the Schrödinger equation, its generalization to the phase space and relation to the Gibbs distribution, Physica Scripta, 2022, vol. 98, № 1, pp. 015221-1-015221-26
26. Perepelkin E.E., Sadovnikov B.I., Inozemtseva N.G., Tarelkin A.A., A new class of exact solutions of the Schrödinger equation, Continuum Mechanics and Thermodynamics, 2019, vol. 31, pp. 639-667
27. G. W. Hill, On the part of the motion of the lunar perigee which is a function of the mean motions of the sun and moon, Acta Math. 1886, vol. 8: pp. 1-36, DOI: 10.1007/BF02417081
28. Mathieu, E., Mémoire sur Le Mouvement Vibratoire d'une Membrane de forme Elliptique, Journal de Mathématiques Pures et Appliquées, 1868, pp 137–203
29. Perepelkin E.E., Sadovnikov B.I., Inozemtseva N.G., Korepanov A.A., PSI-Moyal equation, ArXiV: 2210.07620, pp. 1-22.
30. Perepelkin E.E., Sadovnikov B.I., Inozemtseva N.G., Burlakov E.V., Afonin P.V., The Wigner function negative value domains and energy function poles of the polynomial oscillator, Physica A: Statistical Mechanics and its Applications, 2022, vol. 598, pp. 127339-1-127339-15
31. Perepelkin E.E., Sadovnikov B.I., Inozemtseva N.G., Burlakov E.V., The Wigner function negative value domains and energy function poles of the harmonic oscillator, Journal of Computational Electronics, Springer Nature, 2021, vol. 20, pp. 2148-2158.
32. Perepelkin E. E., Sadovnikov B. I., Inozemtseva N. G. The quantum mechanics of high-order kinematic values //Annals of Physics. – 2019. – vol. 401. – pp. 59-90.
33. Perepelkin E.E., Sadovnikov B.I., Inozemtseva N.G., Burlakov E.V., Explicit form for the kernel operator matrix elements in eigenfunction basis of harmonic oscillator, Journal of Statistical Mechanics: Theory and Experiment, 2020, vol. 2020, № 023109, pp. 1-17
34. Hudson R.L., When is the Wigner quasi-probability density non-negative?, Rep. Math. Phys. 1974, № 6, pp. 240–252.
35. C. M. A. Dantas, I. A. Pedrosa, and B. Baseia, Phys. Rev. A, 1992, 45, 1320
36. J. Y. Ji, J. K. Kim, and S. P. Kim, Phys. Rev. A, 1995, 51, 4268
37. I.A. Pedrosa, Exact wave functions of a harmonic oscillator with time-dependent mass and frequency, Phys. Rev. A, 1997, 55, №4, pp. 3219-3221